\documentclass[useAMS,usenatbib]{mn2e}

\usepackage{dcolumn}
\newcolumntype{d}[1]{D{.}{.}{#1}}

\def\aap{AA}
\def\aapr{AA Rev}

\def\araa{ARA{\&}A}
\def\mnras{MNRAS}
\def\apj{ApJ}

\def\aaps{ApJS}
\def\aj{AJ}
\def\jcap{JCAP}
\def\pasp{PASP}

\def\memsai{MmSAI}

\def\ase{{\prime\prime}}

\usepackage{graphicx}
\usepackage{float}
\usepackage{amssymb}
\usepackage{amsfonts}
\usepackage{amsmath} 
\usepackage{color}

\def\ucla{Department of Physics and Astronomy, PAB, 430 Portola Plaza, Box 951547, Los Angeles, CA 90095-1547, USA}

\def\eso{European Southern Observatory, Karl-Schwarzschild-Strasse 2, 85748 Garching bei M{\"u}nchen, DE}
\def\dark{Dark Cosmology Centre, Niels Bohr Institute, University of Copenhagen, Juliane Maries Vej 30, DK-2100 Copenhagen, Denmark}
\def\unimi{Dipartimento di Fisica, Universit{\`a} degli Studi di Milano, via Celoria 16, I-20133 Milano, Italy}
\def\staples{Staples High School, Westport, CT 06880, USA}
\def\mit{MIT Kavli Institute for Astrophysics and Space Research, Cambridge, MA 02139, USA}
\def\valpo{Instituto de Física y Astronomía, Universidad de Valparaíso, Avda. Gran Breta{\~n}a 1111, Playa Ancha, Valparaíso 2360102, Chile }
\def\abello{Departamento de Ciencias Fisicas, Universidad Andres Bello Fernandez Concha 700, Las Condes, Santiago, Chile}
\def\davis{Department of Physics, University of California, Davis, CA 95616, USA}
\def\subaru{Subaru Telescope, Hilo, HI 96720, USA}
\def\durham{Durham University, Durham DH1 3LE, England}
\def\mil{Millennium Institute of Astrophysics, Chile}
\def\aaemail{\tt aagnello@eso.org}

\title[Quasar lenses and pairs in VST-ATLAS and Gaia]{Quasar lenses and pairs in the VST-ATLAS and Gaia} 
\author[Agnello et al.]{
  A. Agnello$^{1}$\thanks{\aaemail},
  P.~L. Schechter$^{2}$, N.~D. Morgan$^{3},$ T. Treu$^{4,\dag},$ C. Grillo$^{5,6},$\and D. Malesani$^{6},$ T. Anguita$^{7,8},$ Y. Apostolovski$^{7},$ C.~E. Rusu$^{9,10,\ddag},$ V. Motta$^{11},$\and K. Rojas$^{11},$
   B. Chehade$^{12},$ T. Shanks$^{12}$
  \medskip\\
  $^1$\eso\\
  $^2$\mit\\
  $^3$\staples\\
  $^4$\ucla\\
  $^5$\unimi\\
  $^6$\dark\\
  $^7$\abello\\
  $^8$\mil\\
  $^9$\davis\\
  $^{10}$\subaru\\
  $^{11}$\valpo\\
  $^{12}$\durham\\
  $^\dag$ Packard Fellow.\\
  $^\ddag$ Subaru Fellow.\\
  \textit{This paper includes data gathered with the 6.5 meter Magellan Telescopes located at Las Campanas Observatory, Chile.}\\
}
\begin{document}

\voffset-.6in

\date{Accepted . Received }

\pagerange{\pageref{firstpage}--\pageref{lastpage}} 

\maketitle

\label{firstpage}

\begin{abstract}
We report on discovery results from a quasar lens search in the ATLAS public footprint, extending quasar lens searches to a regime without $u-$band or fiber-spectroscopic information, using a combination of data mining techniques on multi-band catalog magnitudes and image-cutout modelling. Spectroscopic follow-up campaigns, conducted at the 2.6m Nordic Optical Telescope (La Palma) and 3.6m New Technology Telescope (La Silla) in 2016, yielded seven pairs of quasars exhibiting the same lines at the same redshift and monotonic flux-ratios with wavelength (hereafter NIQs, Nearly Identical Quasar pairs). The quasar redshifts range between $\approx1.2$ and $\approx 2.7;$ contaminants are typically pairs of bright blue stars, quasar-star alignments along the line of sight, and narrow-line galaxies at $0.3<z<0.7.$ Magellan data of  A0140-1152  (01$^h$40$^m$03.0$^s$-11$^d$52$^m$19.0$^s$, $z_{s}=1.807$) confirm it as a lens with deflector at $z_{l}=0.277$ and Einstein radius $\theta_{\rm E}=(0.73\pm0.02)^\ase$. We show the use of spatial resolution from the Gaia mission to select lenses and list additional systems from a WISE-Gaia-ATLAS search, yielding three additional lenses (02$^h$35$^m$27.4$^s$-24$^d$33$^m$13.2$^s$, 02$^h$59$^m$33.$^s$-23$^d$38$^m$01.8$^s$, 01$^h$46$^m$32.9$^s$-11$^d$33$^m$39.0$^s$). The overall sample consists of 11 lenses/NIQs, plus three lenses known before 2016, over the ATLAS-DR3 footprint ($\approx3500$~deg$^2$). Finally, we discuss future prospects for objective classification of pair/NIQ/contaminant spectra.

\end{abstract}

\section{Introduction}

In an era of large data stemming from ever more ambitious surveys, large samples of intrinsically rare objects become possible. Quasar pairs and strongly lensed quasars are particularly interesting classes of rare astronomical objects, because the information content for each system is high and in some sense unique. Through the lensing effect, one can obtain\footnote{A general review is given by \citet{tre10}.}: (\textsc{i}) a purely gravitational measurement of the properties of the deflector(s), including their invisible components like dark matter halos and individual stars \citep[e.g.][]{ogu14}; (\textsc{ii}) a magnified view of the background quasar, accretion disk, and host galaxy \citep{pen06,slu15,mot17,din17}; (\textsc{iii}) information about distances and thus cosmological parameters \citep{ref64,car13,tre16,suy17}. Spectroscopy of close sightlines, be they to multiple quasar images or to pairs of physically distinct quasars, is a probe of: (\textsc{i}) kinematics of the cosmic web at high redshift \citep{rau05}; (\textsc{ii}) small-scale structure of Ly$\alpha$ absorbers \citep{sme92,din98,ror17}; (\textsc{iii}) physical conditions of the cool ISM/CGM of galaxies and quasars \citep{far14,zah16}.

Unfortunately, lensed quasars are rare on the sky -- typically 1 per 10 square degrees (deg$^2$) at depth and resolution of present day surveys (Oguri \& Marshall 2010) -- since they require a very close aligment of quasars with foreground massive galaxies, or galaxy clusters. Finding them is thus a classic needle in a haystack problem, that requires sophisticated algorithms to identify promising candidates for further follow-up and confirmation. 
In the case of current wide field imaging surveys consisting of thousands of $deg^2$, the data mining problem consists of identifying of order $\sim100$ candidates from catalogs consisting of $10^7-10^8$ astronomical sources.  

In imaging surveys, lensed quasars can be recognized from their colours and morphology. In photometric catalogues, they can appear as quasars with contributions from the lensing galaxy, or galaxies with contributions from the background source, or anything intermediate.  Their image cutouts have morphologies that may be more or less marked, ranging from wide-separation lenses with a clearly visible deflector, to lenses whose image-separation can be ascertained just by direct modelling of the cutouts. In order to ensure maximum efficiency and purity, search strategies have typically been tailored to the specifics of each surveys in the past. Partially-overlapping areas of the Southern Hemisphere are being probed in (at least) $griz$ bands by the DES \citep{san10,des16}, KiDS \citep{dej13}, Pan-STARRS \citep{cha16} and VST-ATLAS \citep{sha15}. The typical resolution of these surveys is just below $1^\ase$ FWHM, in most cases insufficient to fully deblend the multiple images and the deflector for galaxy-scale lenses, at least at pipeline and object-detection level.

Therefore, follow-up is almost always necessary, consisting of spectroscopy to confirm that the quasars are indeed multiply imaged and not a chance alignment, and possibly detect spectral features from the deflector, and high resolution images to map out the lensing configuration. Sometimes confirmation is pretty straightforward, other times it requires substantial observational resources, including Hubble Space Telescope imaging. For example, veritable lenses, like HE1104 \citep{wis93} or HS2209 \citep{hag99}, were identified as `bright' quasar pairs with almost identical spectra and had to wait for deeper and high-resolution follow-up \citep[e.g.][]{cha10} for a full confirmation. Others \citep[e.g. the quad WFI~2026,][]{mor04} are lacking secure spectroscopy of the deflector to this day. These systems are not uncommon in quasar lens searches: the Sloan Quasar Lens Search \citep[SQLS:][]{ogu06,ina12,anu16} yielded some quasar pairs with nearly identical spectra at the same redshifts, but not deflectors detected. In small separation lenses, the deflector may be faint enough to be undetectable unless the quasar images are subtracted from deep, high-resolution imaging data \citep[e.g. for HS2209,][]{wil17b}.

Here, we report on spectroscopic follow-up results from a two-step search applied mainly to the VST-ATLAS public footprint\footnote{Accessible at \texttt{http://osa.roe.ac.uk/}}.
In Section 2, we describe the \textit{target} and \textit{candidate} selection procedures; follow-up campaigns are summarized in Section 3; results are discussed in Section 4 and Table~\ref{tab:bigtab}; we discuss future prospects in Section 5, and list additional candidates (Table~\ref{tab:morecands}) that could not be followed up in 2016, including the first quasar lens candidates identified using Gaia data and three new lenses among them.
Target selection was based on $g,r,i,z$ and WISE \citep{wri10} $W1,W2$ magnitudes. For consistency with previous work, the WISE magnitudes were left in the Vega system, whereas the ATLAS magnitudes were translated in the AB system.

\section{Targets and Candidates}

The strategy followed here consists of two steps. First, \textit{targets} are selected from the ATLAS catalogs based on their magnitudes in optical and infra-red magnitudes. Then, \textit{candidates} are obtained by retaining just the targets that pass a first visual inspection and modelling their cutouts, to ensure that they are consistent with two or more point-sources with consistent colours.
A similar search was performed on a patch of the SDSS footprint with right ascensions accessible to observation around February 2016, further concentrating on four objects with SDSS fibre spectra that were used as a control sample.

A different kind of target mining, based on outlier selection \citep{agn17}, was applied to the ATLAS-DR3 footprint once it became publicly available in November 2016. For the ATLAS-DR3 and SDSS targets, the only candidate selection step consisted in visual inspection.

\begin{figure}
 \centering
 \includegraphics[width=0.23\textwidth]{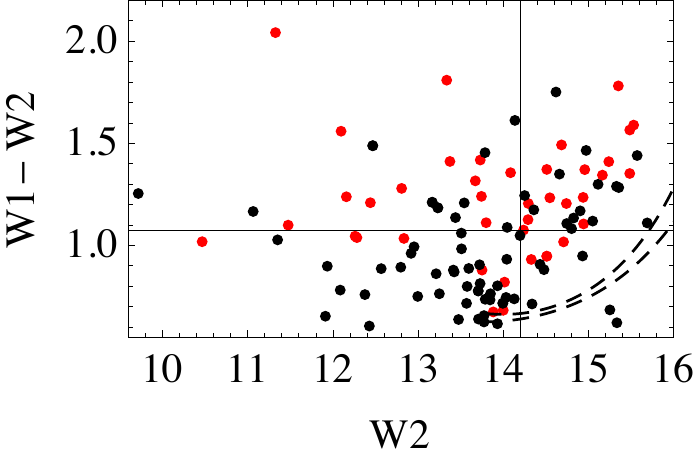}
 \includegraphics[width=0.23\textwidth]{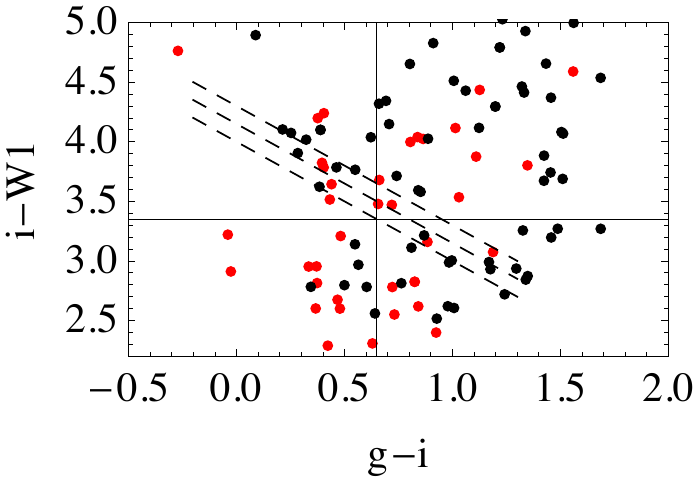}\\
 \includegraphics[width=0.225\textwidth]{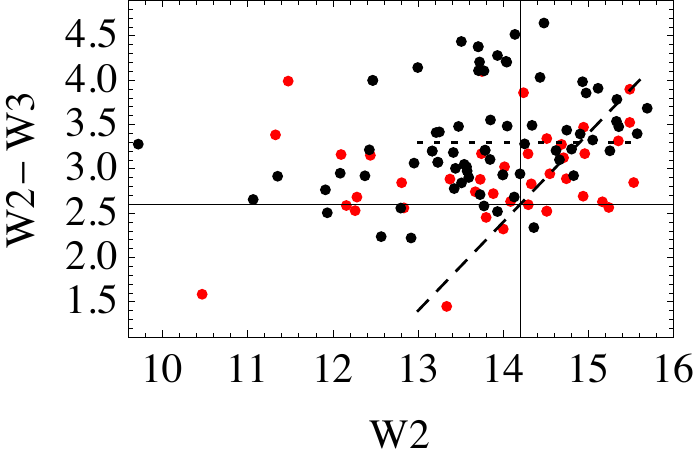}\hspace{0.005\textwidth}
 \includegraphics[width=0.225\textwidth]{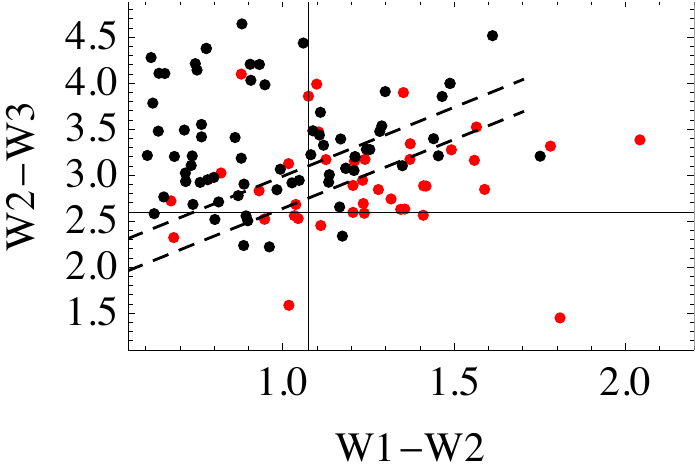}\\
  \includegraphics[width=0.45\textwidth]{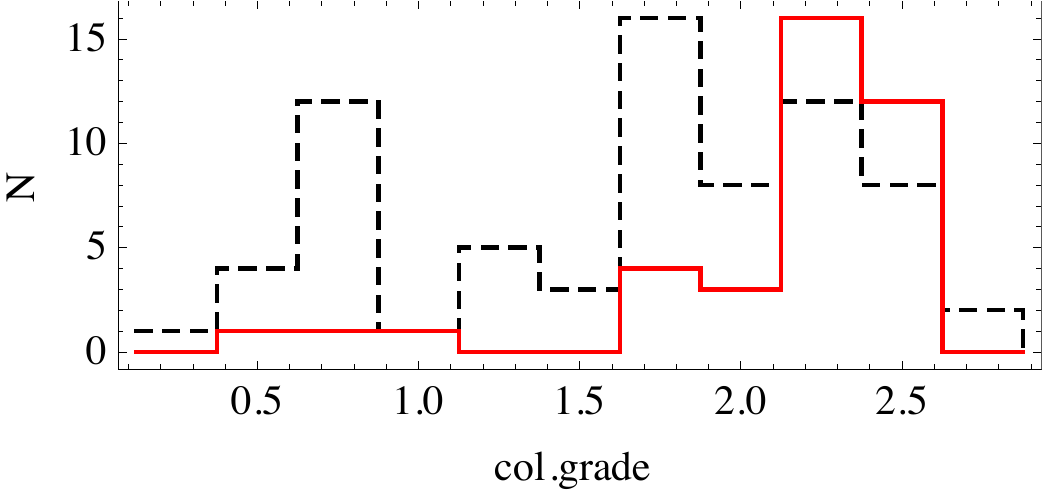}
\caption{{Colour-magnitude diagrams of ANN and IRX targets that passed a first round of visual inspection. The red (resp. black) points indicate candidates that were accepted (resp. rejected) by the model selection of Section~\ref{sect:cands}}. The dashed lines represent the \citet{ass13} locus (top-left panel), the detection limit $W3=11.6$ (middle-left panel), a rough separation of  high-$z$ quasars from lower-$z$ quasars and early-type galaxies (top-right) and galaxy-like and lens-like objects (middle-right). The bottom panel shows the colour-grade histograms of model-accepted (resp. rejected) candidates in red (resp. black dashed).}
\label{fig:targprop}
\end{figure}
\subsection{ATLAS DR2 Target Selection}
\label{sect:tarsel}

The coverage of ATLAS over its footprint is not uniform in all bands. Then, different selection procedures were adopted for different combinations of bands, in order to maximize the target sample. When querying objects from the ATLAS public footprint, we required an extendedness criterion, given by either \texttt{p{\_}Galaxy}$>0.5$ or \texttt{AperMag3{\_}i-AperMag6{\_}i}$>0.08,$ i.e. that the objects have extra flux besides that of an isolated point-source.

\subsubsection{Artificial Neural Networks}
When $griz$ bands were available, we selected objects that are `blue and extended', using colour cuts along the lines of \citet{agn15b}, but without restrictions on $i-W2$ or $g-i.$ This is to avoid the exclusion of higher-redshift quasars ($z_{s}\gtrsim 2$) and redder objects where a lensing galaxy could contribute more to the colours. For these objects, we used Artificial Neural Networks (ANNs) to select those that were compatible with lensed quasars, or quasars at redshift $z_{s}>0.75;$ this was made possible by extending the ANNs designed by \citet{agn15} to split the `quasar' class into multiple redshift intervals \citep[cf][]{wil16}, bringing the total number of classes to nine from the initial four that were used by \citet{agn15b}.

\subsubsection{Missing magnitudes and hybrid colours}
In the absence of some optical bands, we could still select some objects based on their infra-red excess, i.e. optical colours resembling those of quasars and redder optical-infrared colours that could be indicative of a lensing galaxy. This approach was used successfully by \citet{ofe07} in the case of the SDSS, and we used it here with either $g-r$ or $r-z$ for the optical colours and $r-H$ or $r-K_{s}$ for the hybrid colours, whenever $H$ or $K_{s}$ magnitudes are available from 2MASS \citep{skr06}.

An additional sub-sample of targets consisted of objects that satisfied some strict colour-magnitude cuts
\begin{eqnarray}
\nonumber i-W1<3.7,\ \ g-i<0.65,\\
 \ W1-W2>1.075,\ W2<13.4\ .
\end{eqnarray}
This identifies the locus where 7 out of the 10 small-separation lenses of \citet{ina12} lie. They are dominated by the source quasar, being blue in the optical and having a high WISE excess, and a low $i-W1$ typical of quasars at higher redshift.

\subsubsection{Colour grading of DR2 targets}
Targets in the ATLAS DR2 footprint were also graded based on their colours, where a grade of 0 (resp. 3) means low (resp. high) chances to be a quasar lens. The colour grade was assigned as $s=1+s_{1}+s_{2}+s_{3}+s_{4},$ with
\begin{eqnarray}
\nonumber s_{1}=\theta(-0.5(g-r+0.8(u-g-0.6)-0.4))\\
\nonumber s_{2}=0.5H(3.6-(W2-W3))-0.5H(W2-W3-3.6)\\
\nonumber s_{3}=0.5H(3.1+1.5(W1-W2-1.075)-(W2-W3))\\
s_{4}=0.5H(3.4-(i-W1)),
\end{eqnarray}
where the Heaviside step function $H(x)$ is 1 (resp. 0) for $x>0$ (resp. $x<0$), and $\theta(x)=x$ for $-0.5<x<0.5$ and  $\theta(x)=0.5$ (resp $-0.5$) for $x>0.5$ (resp. $<-0.5$). Whenever a magnitude is not available (especially $u$), the grade to which it contributes is set to 0.

\begin{figure*}
 \centering
 \includegraphics[width=0.33\textwidth]{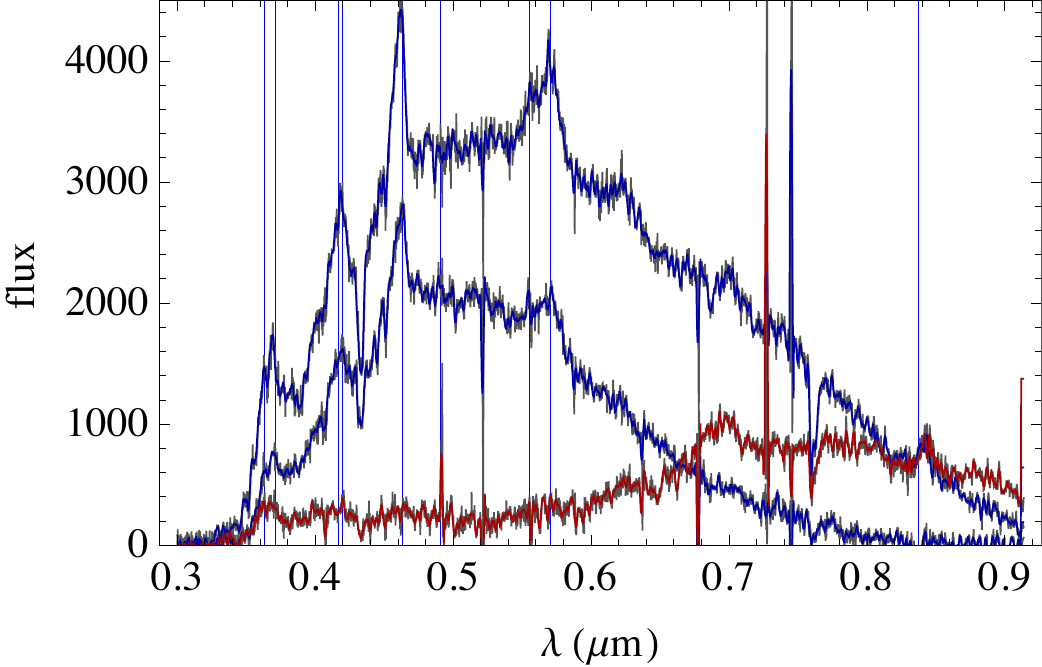}
 \includegraphics[width=0.33\textwidth]{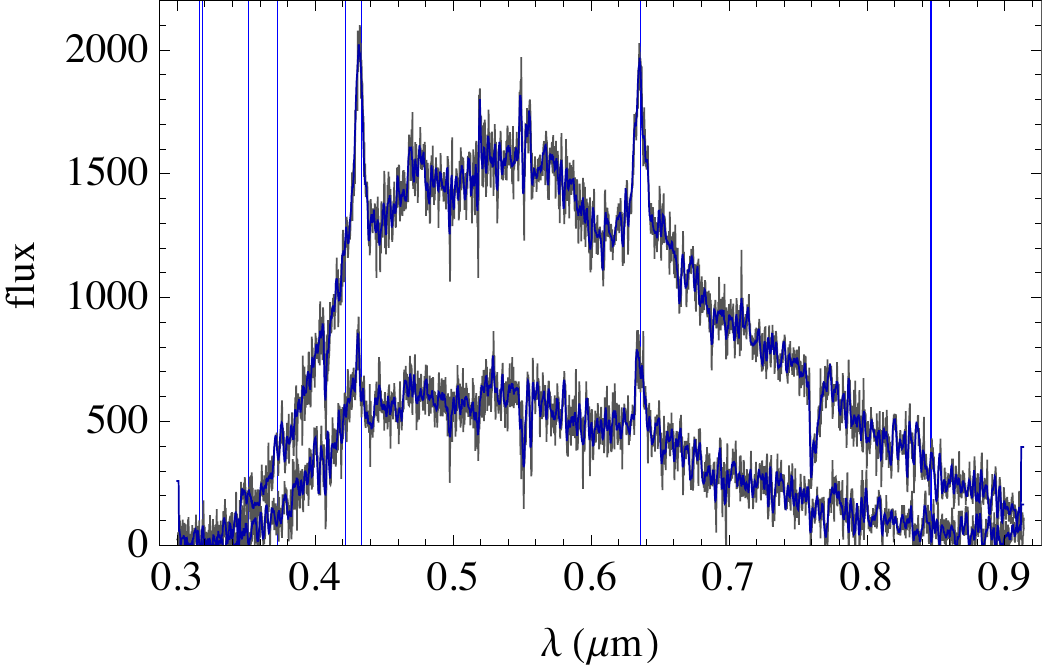}
 \includegraphics[width=0.33\textwidth]{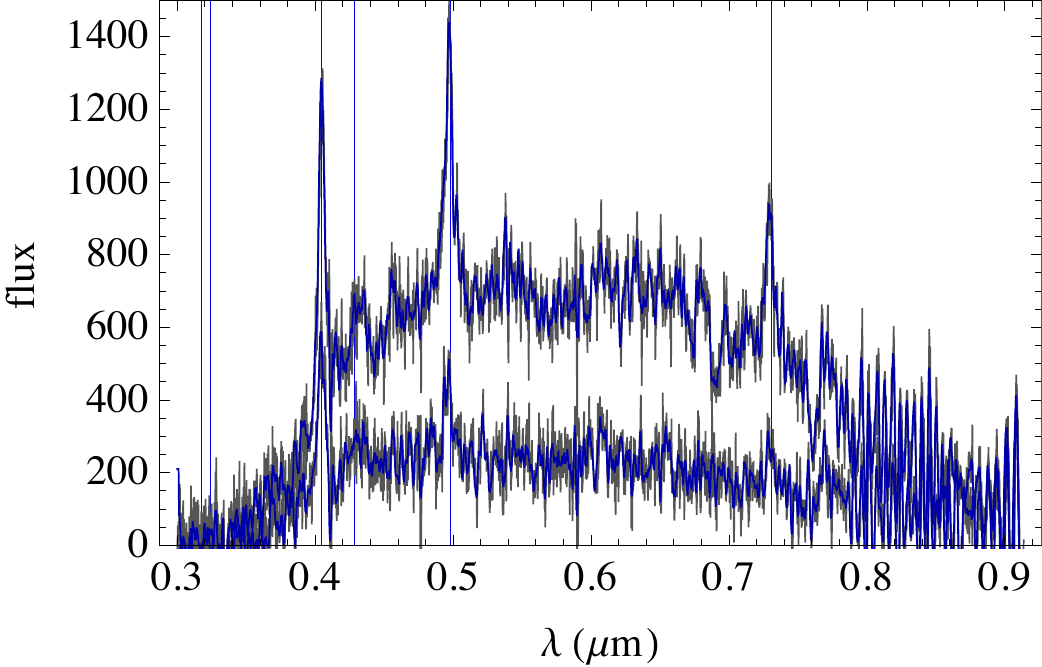}\\
 \includegraphics[width=0.33\textwidth]{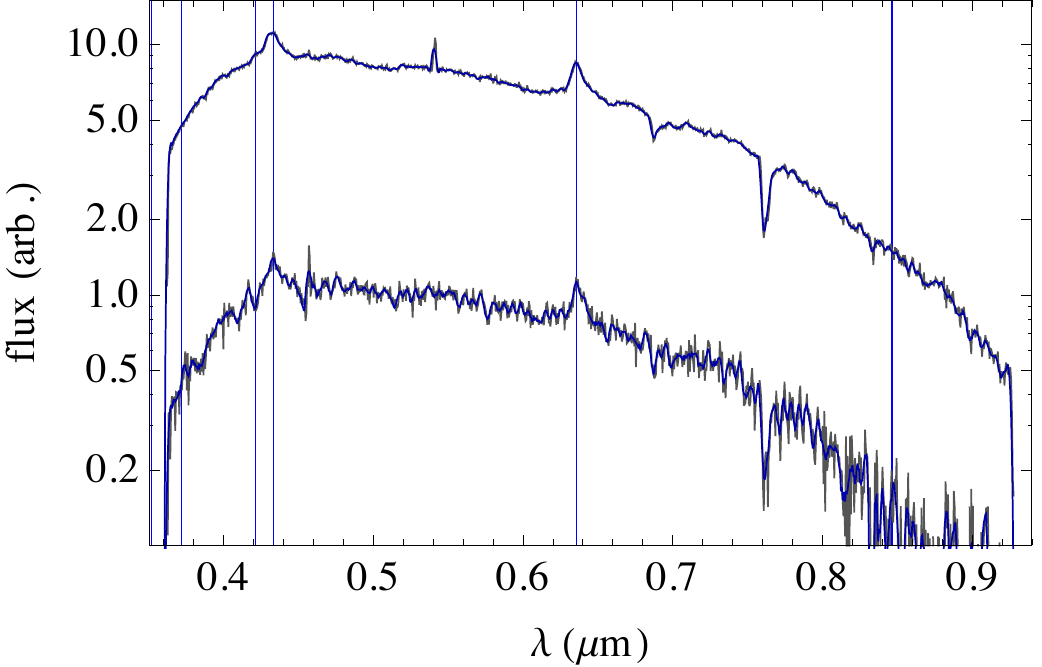}
 \includegraphics[width=0.33\textwidth]{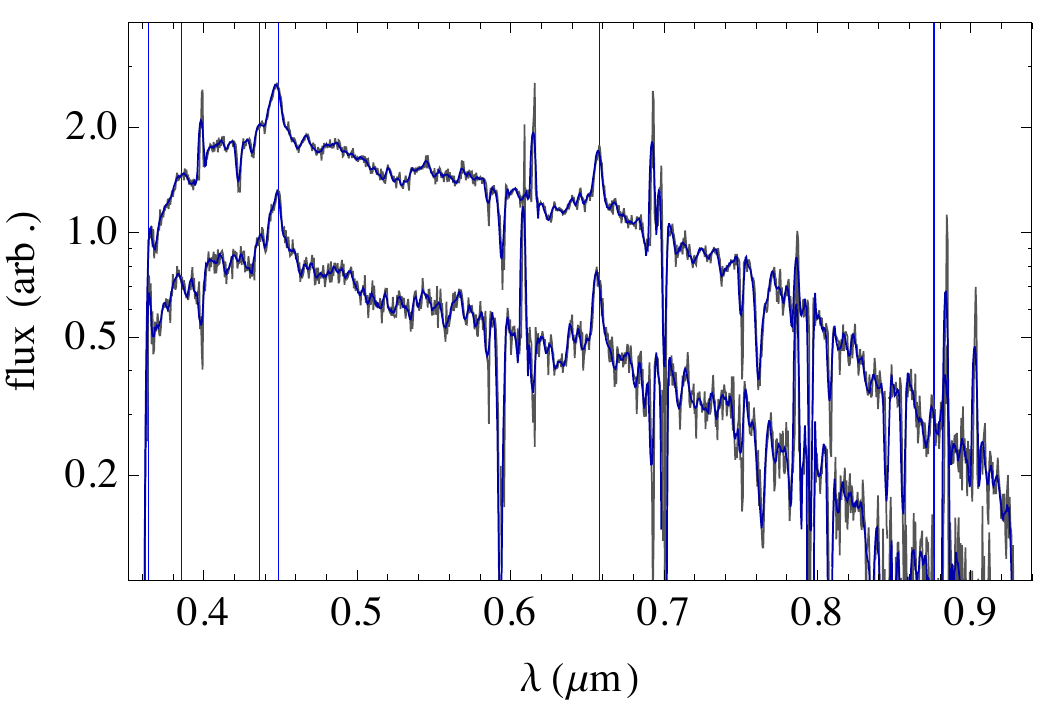}
 \includegraphics[width=0.33\textwidth]{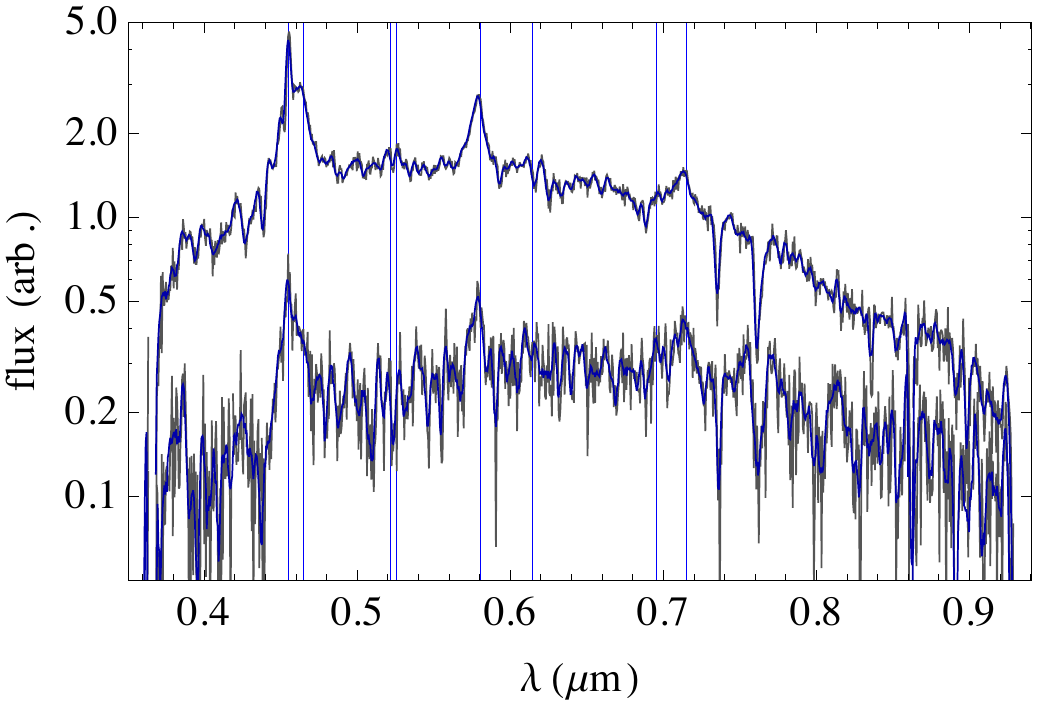}
\caption{{Deconvolved long-slit spectra of six NIQs (see table \ref{tab:bigtab}). \textit{Top:} NOT spectra of A1132-0730, A1112-0335 and S1128+2402, with $z_{s}=1.99,$ $1.27$ and $1.608$ respectively. The estimated $z_s$ for S1128 agrees with the one from SDSS fibre spectra; the wavelength calibration is probably inaccurate towards the red end. A1132 shows broad Fe\textsc{ii} emission (too weak in the other spectra) and a red excess to the opposite side of the fainter image, whose origin is uncertain. \textit{Bottom:} NTT spectra of A2213-2652, A0326-3122 and A1012-0307, with $z_{s}=1.27,\ 1.35$ and $2.745$ respectively. A1012-0307 is the known lens LBQS~1009-0252 \citep{hew94}, blindly rediscovered and shown here for a discussion of NIQ/pair classification.
  }}
\label{fig:confspec}
\end{figure*}

\subsection{Candidate Selection}
\label{sect:cands}

From the previous step, we obtained a pool of targets that were further refined to obtain a final candidate sample. First, the targets were visually inspected by three of us (AA, TT, CER) to exclude obvious contaminants, such as galaxies, low-redshift quasars with a bright host, line-of-sight quasar-galaxy alignments, isolated objects and pairs with colours that were manifestly inconsistent. As a second step, the multi-band ATLAS-DR2 cutouts were modelled automatically to verify whether the objects could be `split' into two (or more) point-sources with consistent colours across the available magnitudes, as described below.

\subsubsection{Candidate Corroboration}
Fourteen of the 15 DR1 and DR2 candidates for which spectra were eventually obtained were independently evaluated as candidates using the ATLAS cutout morphology approach described by \citet{sch16}. Of these, ten had been independently targeted by them for cutout evaluation based solely on a simple cut on the $W1-W2$ colour.

Besides accepting or rejecting a target, this procedure also assigned grades corresponding to different diagnostics \citep[see][for details]{sch16}. For the sake of completeness, we retain the \texttt{ufom} overall figure of merit in what follows, even though it was not used to prioritize candidates for follow-up.

\subsubsection{General candidate properties}
The cutout-modelling stage enabled a further refinement of the visual-inspection survivors into model-accepted and model-rejected candidates. Their selected colour-magnitude diagrams are displayed in figure \ref{fig:targprop}.
The histograms of colour-grade of the model-accepted and rejected targets are shown in the bottom panel. Between half and two thirds of the DR2 targets with grade$>2$ are accepted by the cutout modelling.

From the colour-magnitude diagrams, model-rejected targets tend to lie at lower $W1-W2$ or higher $W2-W3,$ which are regions typically populated by galaxies or, at best, quasars with an extended host. The dashed lines in $W1-W2$ vs $W2-W3$ separate most of the model-accepted candidates from the rest, with different thresholds. The upper line also happens to separate the 10 SQLS small separation lenses of \citet{ina12} from about half of the 40 false-positives in that search. This suggests that $W2-W3$ adds information over the original ANN implementation \citep[see in particular][for a detailed discussion]{wil16}, and in fact this is used in the outlier selection method \citep{agn17} that has been applied to the DR3 data.


\section{Follow-up}
Long-slit spectroscopy was used to ascertain the nature of the candidates. The DR2 and SDSS candidates were mostly observed with the Andalucia Faint Object Spectrograph and Camera (ALFOSC) at the 2.6~m Nordic Optical Telescope (NOT) in La Palma (Spain). The remaining candidates from DR2 and DR3 were observed with the ESO Faint Object Spectrograph and Camera (EFOSC2) at the 3.6~m New Technology Telescope (NTT) in La Silla (Chile). The objects are listed in Table~\ref{tab:bigtab}.

For one object, A0140-1152, we also show spectra taken with IMACS at the 6.5m Walter Baade Telescope at Magellan (Las Campanas), where a red galaxy at $z_{l}=0.277$ is detected between two quasar images at $z_{s}=1.807,$ making this system most likely a lens. This system was observed on UT Nov 28, 2016, IMACS was set up in `long' f/4 camera mode, and a $3800-7000\rm{\AA}$ filter was used.

Custom routines were used for data reduction, as specified below for the two setups. The sky subtraction and 1D spectral extraction were performed in a second stage, on the reduced 2D spectra \citep{agn15b, sch16}. At each wavelength pixel, the raw data in the spatial direction were modeled as a superposition of two (or more) Gaussians and a spatially-uniform component for the sky lines. Within each science frame, the Gaussians are forced to have the same FWHM and the same separation across the whole wavelength  range, even though their peak positions are allowed to vary (linearly) with wavelength. The extracted 1D spectra of each component were then co-added, and data and noise spectra were obtained via de-trended fluctuation analysis, with a 5+5 wavelength-pixel window and a quadratic polynomial for the de-trending.
This procedure allowed us to reliably separate the multiple components in all observed candidates despite sub-optimal observing conditions, with seeing FWHM between $\approx 1^\ase$ and $1.5^\ase$ and clouds during the NOT observations, and variable weather and seeing during the NTT observations.

\subsection{NOT Follow-up}

\begin{figure}
 \centering
 \includegraphics[width=0.235\textwidth]{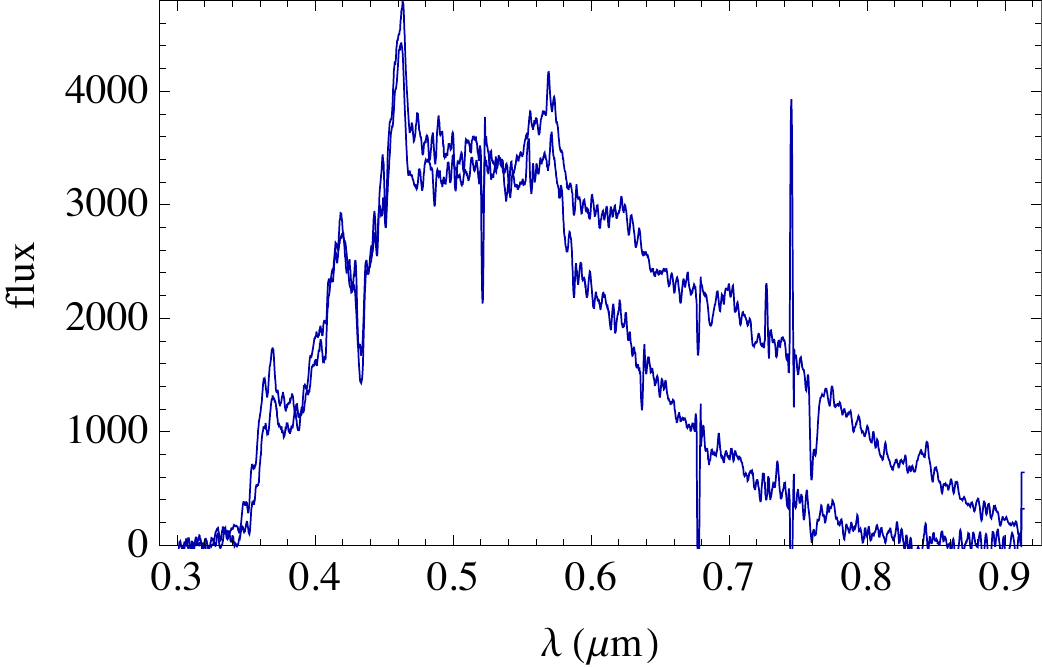}
 \includegraphics[width=0.235\textwidth]{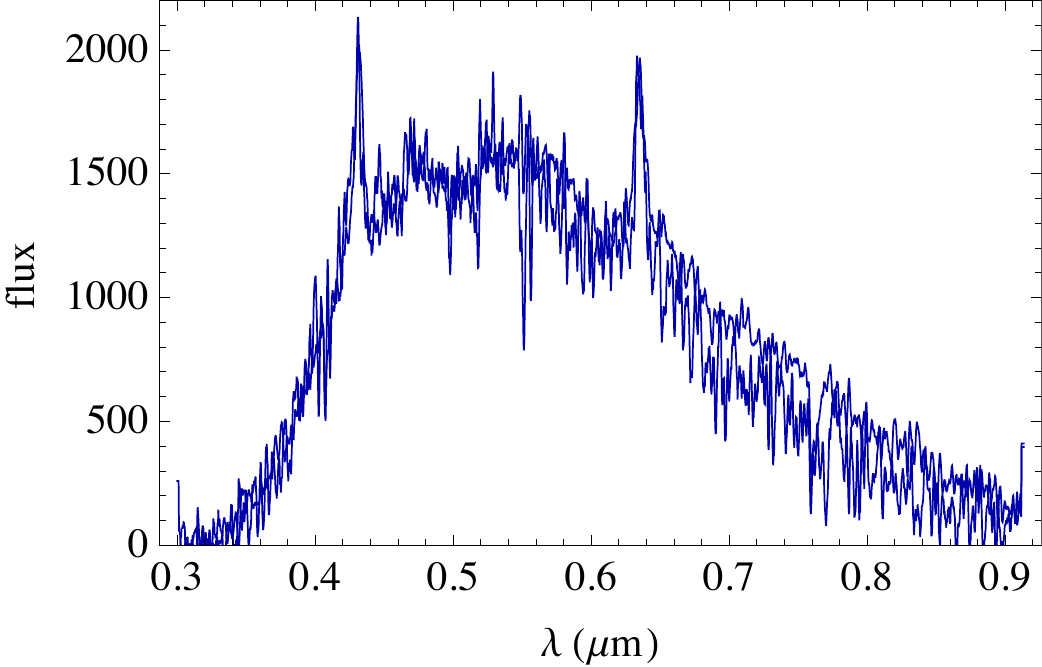}\\
 \includegraphics[width=0.235\textwidth]{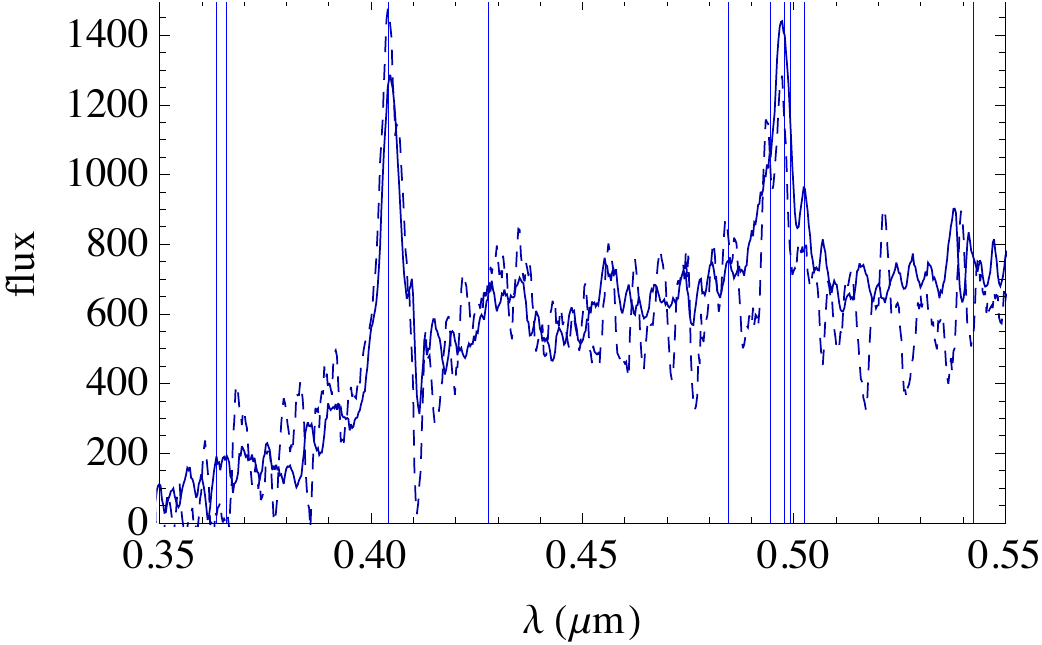}
 \includegraphics[width=0.235\textwidth]{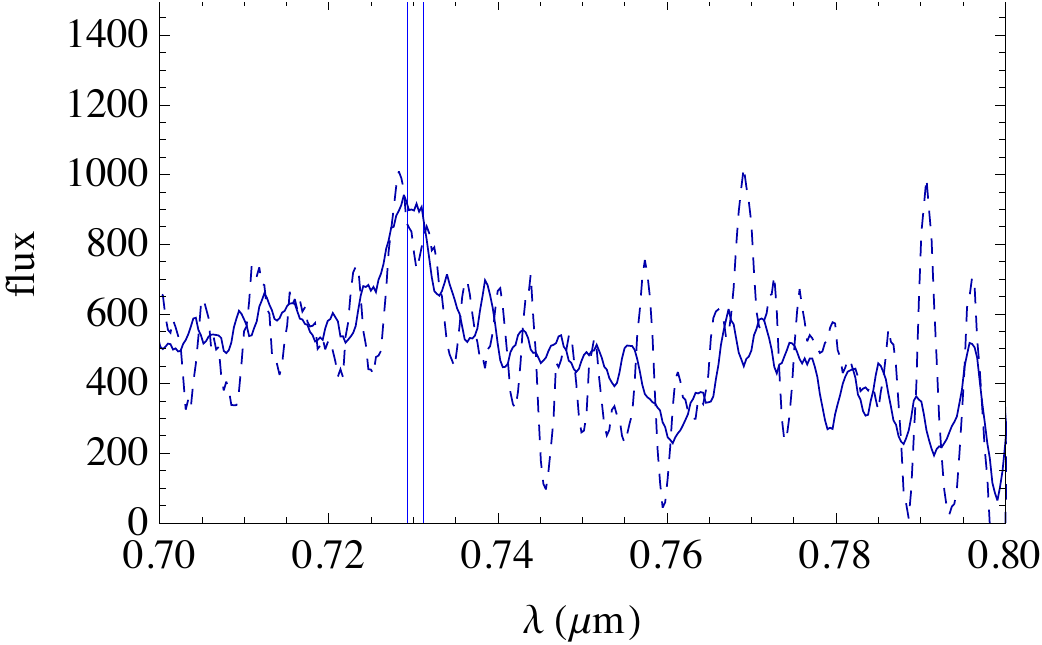}
\caption{{Rescaled fainter-component spectra superimposed on the brighter component, for the NIQs from the campaign at NOT: top-left for A1132 (resc.~1.7), top-right for A1112 (resc.~2.5) and bottom panels for J1128 at the blue and red extrema (resc. 2.5, 3.2). The flux ratios of A2213 and A0326 (not shown here), as measured on the continuum between C\textsc{ii}$\left.\right]$-Fe\textsc{iii} complex and Mg\textsc{ii}, are 8.25 and 2.3 respectively. A1012 (LBQS1009-0252) has flux ratio well approximated by $3.0(\lambda/9000~\rm{\AA})^{-1}.$}}
\label{fig:specdet}
\end{figure}

The data were obtained on 2016 February 5 and 6 as part of a Niels Bohr Institute Guaranteed Time Observing Program (P52-802). We positioned $1^\ase-$wide long-slits through the candidate multiple images and used ALFOSC with the {\#}4 grism, covering thea wavelength range $3200~\rm{\AA} <\lambda< 9600~\rm{\AA}$ with a dispersion of $3.3~\rm{\AA}/\rm{pixel}$. Two science exposures were taken per object, with arc (HeNe, Ar) and flat lamps bracketing each observing block. Standard \textsc{Iraf} routines were used for bias subtraction, flat-field corrections and wavelength solution. 


\subsection{NTT Follow-up}
The data were obtained on 2016 Sept.25-27th and Dec.5th at the ESO-NTT (PI Anguita, 097.A-0473(A), 098.A-0395(A)). The $1.2^\ase-$wise long-slit in combination with EFOSC grism {\#}13 was used, covering  $3400~\rm{\AA} <\lambda<10000~\rm{\AA}$ with $\approx5.5~\rm{\AA}/\rm{pixel}$. Mostly one exposure was taken per object, with calibrations taken once per night. The ESO-provided pipeline (v2.5.5) was used for data reduction.

%

\begin{figure*}
 \centering
 \includegraphics[width=0.33\textwidth]{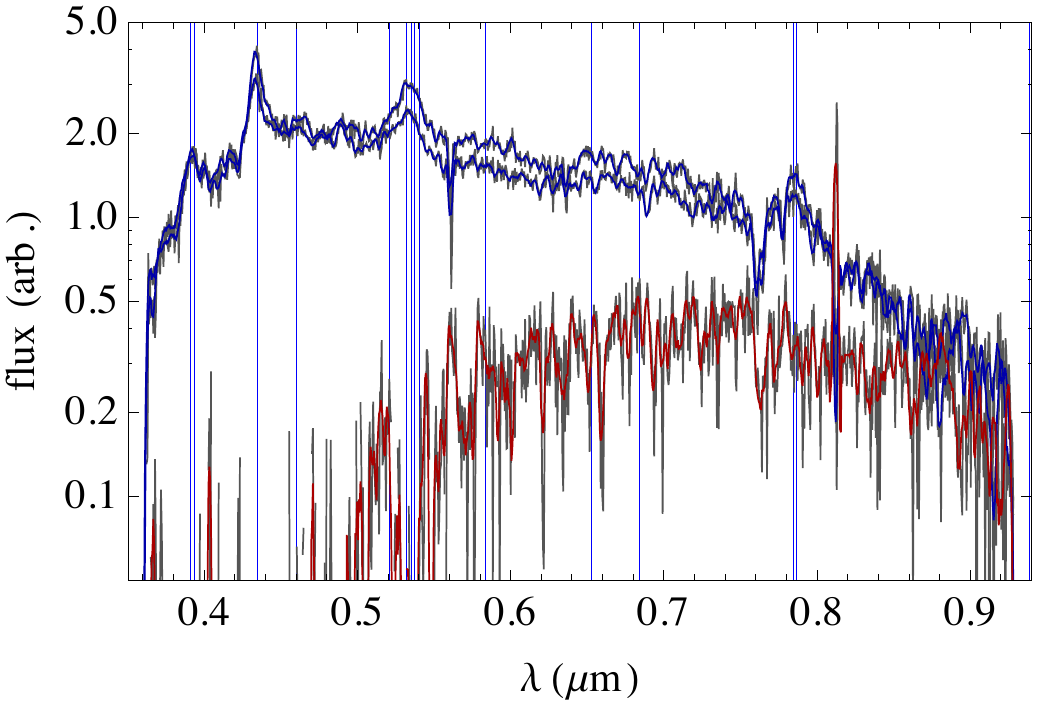}
 \includegraphics[width=0.33\textwidth]{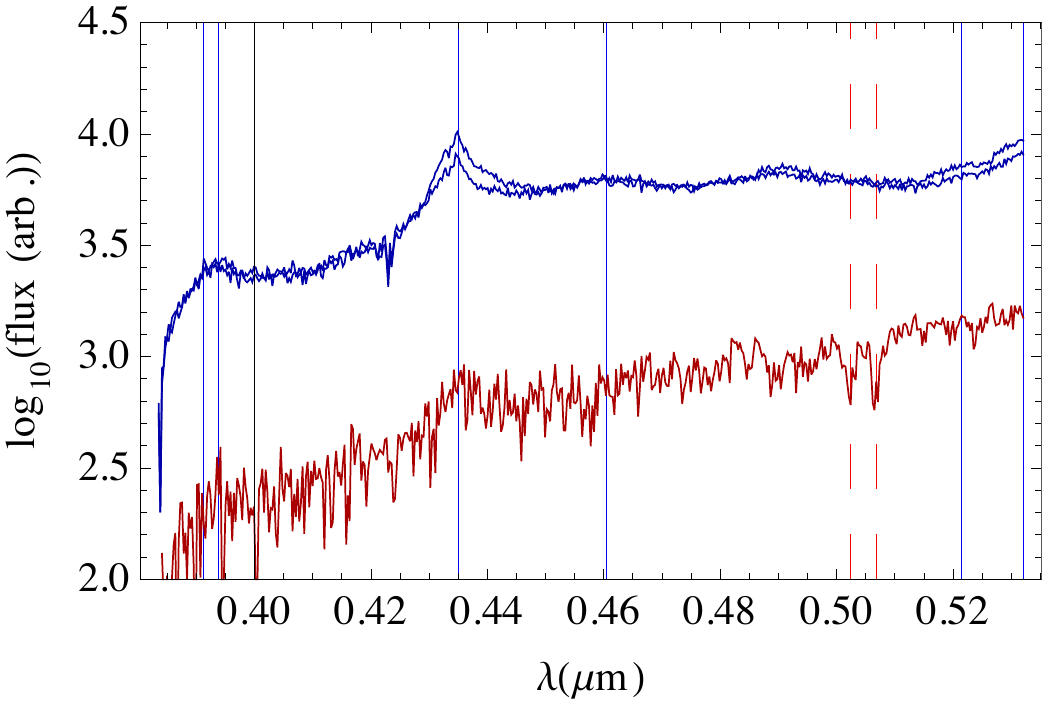}
 \includegraphics[width=0.33\textwidth]{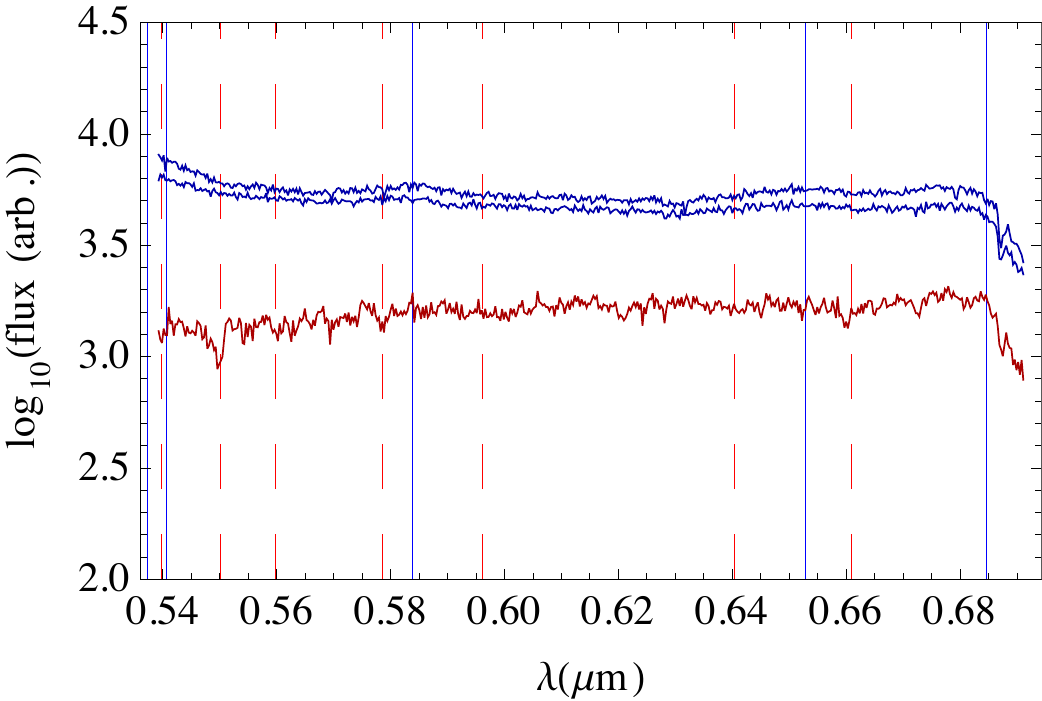}\\
 \includegraphics[width=0.99\textwidth]{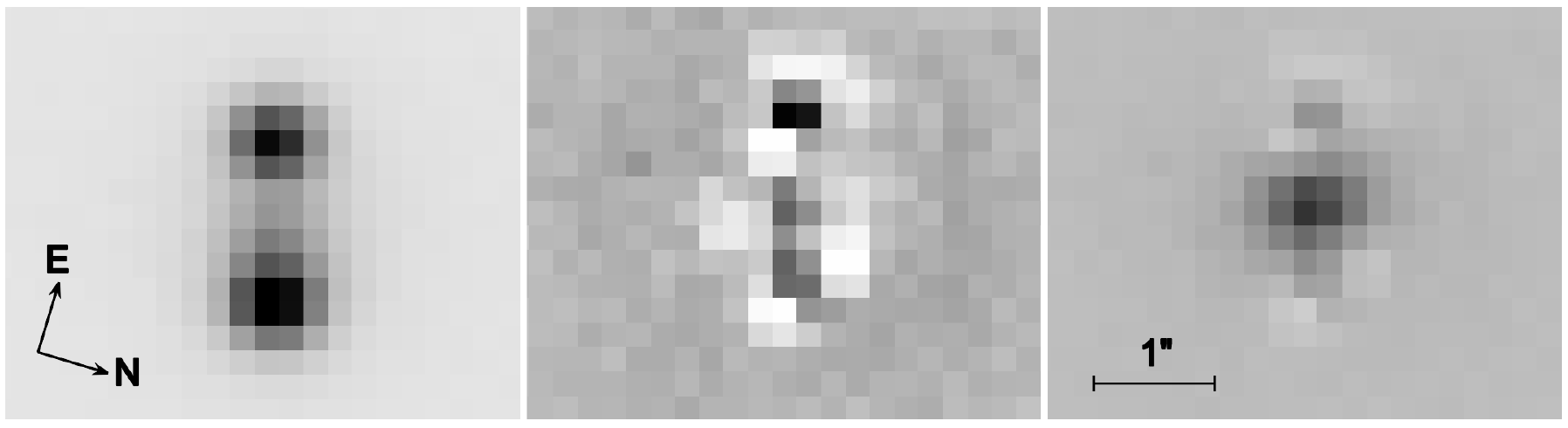}
\caption{{Follow-up data of A0140-1152. \textit{Top left}: EFOSC spectra, showing two almost-identical $z_{s}=1.805$ quasar spectra with flux ratio $f_{r}\approx1.25,$ plus a red excess (located between the quasar images). \textit{Top middle, right}: IMACS spectra, showing identical $z_{s}=1.807$ quasar spectra with $f_{r}\approx1.05$ between C\textsc{iv} and C\textsc{iii}$\left.\right]$ and $f_{r}\approx1.12$ between Fe\textsc{iii} `uv48' and C\textsc{ii}, plus the same red excess with Ca (G,H,K) absorption. \textit{Bottom left:} IMACS $i-$band acqusition image (0.2$^\ase$/px, left), aligned with the slit at p.a.=106.7~deg E of N. \textit{B. middle}: residuals after subtracting lens galaxy and quasar images. \textit{B. right:} same model, but only the quasar images are subtracted.}}
\label{fig:A0140}
\end{figure*}

\begin{figure}
 \centering
 \includegraphics[width=0.235\textwidth]{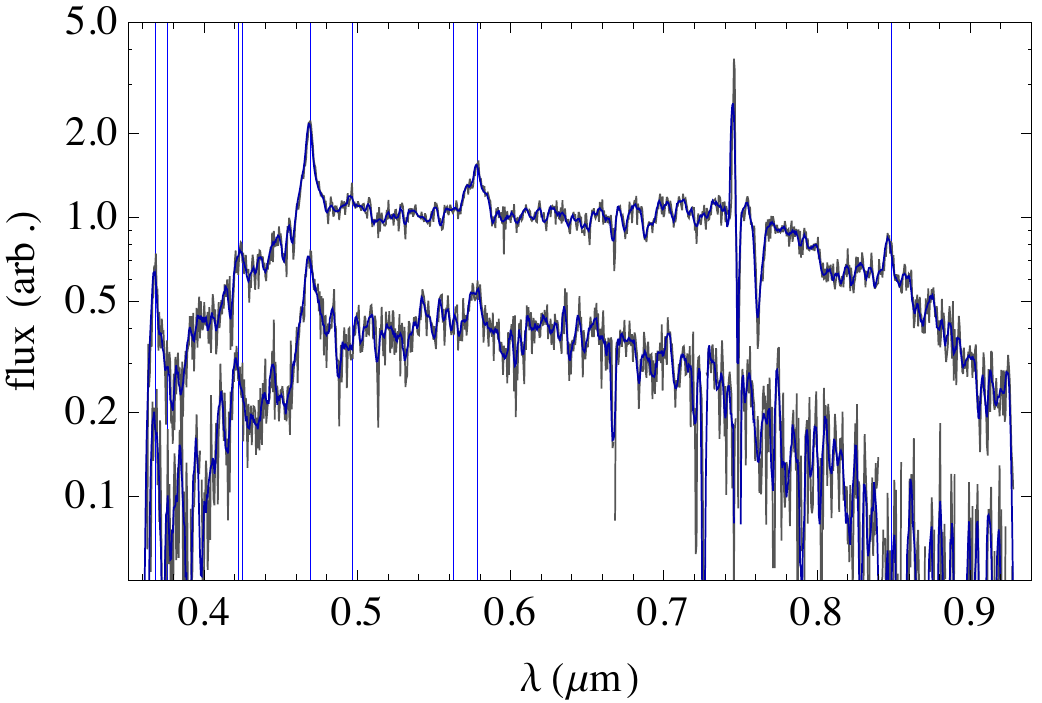}
 \includegraphics[width=0.235\textwidth]{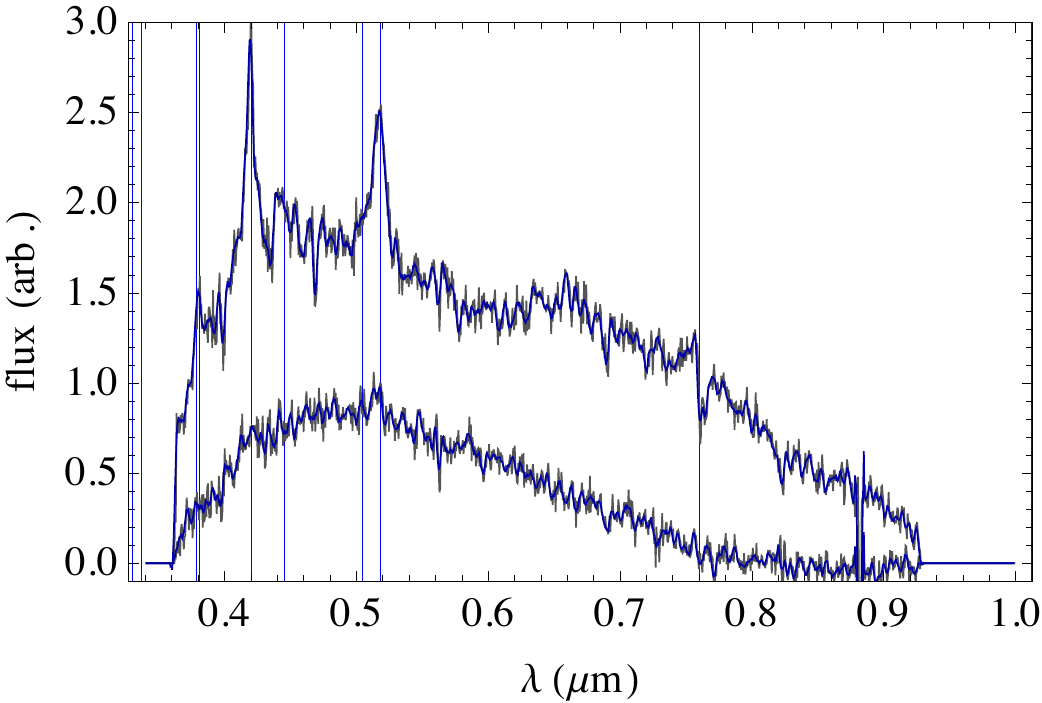}\\
 \includegraphics[width=0.235\textwidth]{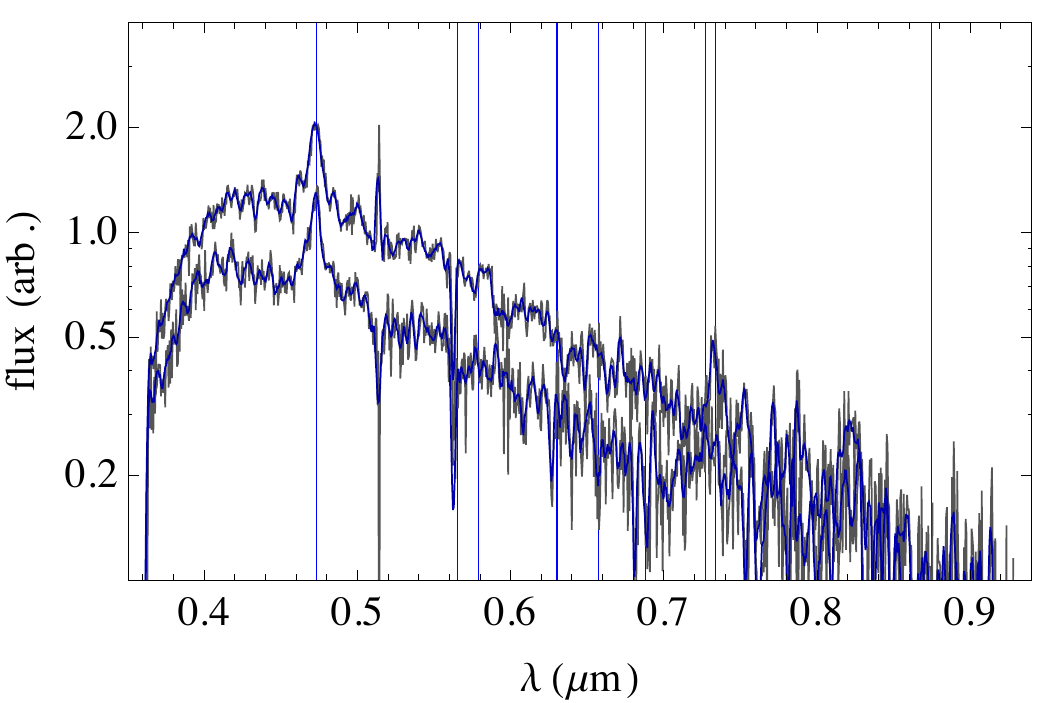}
 \includegraphics[width=0.235\textwidth]{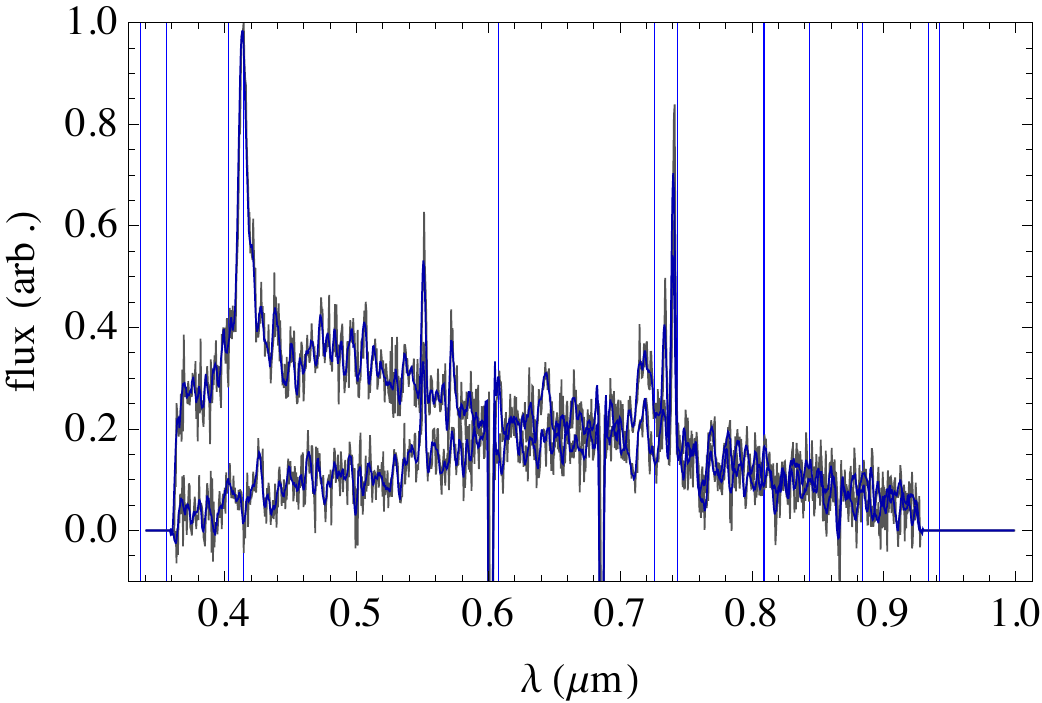}\\
\caption{{Spectra of A1020 ($z=2.03,$ top left), A0008 (top right), A2338 (bottom left), A0054 (bottom right).}}
\label{fig:others}
\end{figure}

\section{Discussion and Conclusions}

Long-slit spectra of twenty-seven objects were obtained (tab.~\ref{tab:bigtab}), out of which seven were \textit{nearly identical quasar pairs} (NIQs), i.e. pairs of quasars with the same lines at the same redshift and smooth flux-ratios as measured on the continua. These could be veritable lenses where the deflector has not been detected yet, or truly physical pairs of quasars. At least one of them, A0140-1152, is a lens based on the deeper Magellan spectra (fig.~\ref{fig:A0140}). An eighth system, A1020-1002, is a pair of quasars at $z=2.03$ (fig.~\ref{fig:others}) and could also be a NIQ, with flux ratio $f_{r}=2.3$ measured between C~\textsc{iii}$\left.\right]$ and Mg~\textsc{ii}, but deeper data are needed. Some, peculiar false positives are discussed in Sect.\ref{sect:pecfalpos}. 

\subsection{Near-Identical Quasar Pairs}
Seven objects have two quasar spectra with the same lines at the same redshift and same shapes, and monotonic flux-ratios as measured on the continua (figs.~\ref{fig:confspec},\ref{fig:specdet},\ref{fig:A0140}). Four, marked by asterisks in Tab.~\ref{tab:bigtab}, were also flagged independently in the search described by \citet[][ whence the terminology of NIQs is adopted]{sch16}. In all cases the traces could be reliably deconvolved, and their flux ratios vary from $\approx1$ to $\approx10$ for the sample. One NIQ was found among the four SDSS targets with quasar fibre spectra, the others being quasar+quasar or quasar+star alignments. Two systems in particular are described below. 

\subsubsection{A1012-0307, a known lens with power-law flux-ratios}
A1012-0307 is the known lens LBQS1009-0252  \citep{hew94}. The NTT data can be well deblended into two $z_{s}=2.745$ quasar spectra, showing the same lines with the same shape (fig.~\ref{fig:confspec}). The redshift from NTT spectra is slightly higher than the one ($z_{s}=2.739$) quoted in its original discovery paper. Differently from the other NIQs, whose flux ratios are almost constant with wavelength, in this case the flux ratio is well fitted over the whole spectral range by $f_{r}=3.0(\lambda/9000\rm{\AA})^{-1}.$ In principle, such a variation could be produced by intrinsically different continua in two physically separate quasars, or significant differential reddening by a foreground galaxy. In simple models of doubly imaged quasars (without shear), the fainter image forms closer to the lens and so can be more heavily reddened. Still within the lensing hypothesis, also microlensing should be considered as a possible source of chromatic flux ratios.

\subsubsection{A0140-1152, a new quasar lens}
\label{sect:A0140}
A0140 was observed at the NTT and Magellan independently (fig.~\ref{fig:A0140}), having been respectively targeted in the ATLAS-DR3 catalog and separately found via cutout modelling in the whole footprint. Both data-sets revealed $z_{s}=1.806\pm0.001$ nearly-identical quasar spectra, plus a red `excess' between the two traces that the Magellan spectra and acquisition images confirmed to be a galaxy at $z_{l}=0.277$ with prominent Ca (G,H,K) absorption features. The flux-ratio between the two quasar images, as measured on the continua, is $f_{r}=1.05$ between C lines on the blue side and $f_{r}=1.12$ on the red side; the NTT spectra gave $f_{r}=1.2$ between the C~\textsc{iii}$\left.\right]$/Fe complex and Mg~\textsc{ii}. The small discrepancy may be ascribed to lower S/N, atmospheric dispersion corrections and slight slit misalignment.

For this system, we also run simple lens models. From the IMACS $i-$band acquisition image, aligned with the slit (fig.~\ref{fig:A0140}), the relative positions and magnitudes of the quasar images and deflector have been obtained, as listed in table~\ref{tab:astromA0140}. The flux ratio between images A and B varies depending on whether it is measured on the continua or on (continuum-subtracted) emission lines, or on the broad-band IMACS image. Besides instrumental effects (chromatic atmospheric refraction and slit-losses), this chromaticity results from differential extinction in the deflector and microlensing. We then adopt $f_{A}/f_{B}=1.20\pm0.05$ as a measured flux-ratio constraint, which is obtained on the emission lines and accounts for systematics in continuum-subtraction, differential extinction, and instrumental effects. The observational constraints are then: the positions of images A,B relative to G; and the flux ratio $f_{A}/f_{B}.$ An additional constraint is the presence of only two (observed) quasar images.

The deflector (G) is described as a Singular Isothermal Ellipsoid \citep[SIE,][]{kas93}, and we consider an external shear component to account for (possible) corrections to the quadrupole, e.g. from additional mass along the line of sight or by deviations from a simple, SIE model. 
A model with free shear and ellipticity is unconstrained, so we explore three restricted models: mod.(a, SIE) adopts a SIE for the deflector, without external shear; mod.(b, SIS+XS) adopts the spherical limit ($q\rightarrow$1) for G and includes external shear; mod.(c,SIE+XS) includes external shear, and has G described by a SIE with $q=0.5$ and p.a.$=28.0$~deg (E of N) as inferred from the IMACS image.

A model with SIE+XS and wider, uniform priors on all parameters is discussed in the Appendix. Since this last model is unconstrained, we use it to characterize the quadrupole degeneracy between shear and ellipticity. To this aim, we adopt unrealistically low positional uncertainties, $\delta x=\delta y=0.01^{\ase},$ exploring a `tube' of solutions. 

\begin{figure}
 \centering
 \includegraphics[width=0.45\textwidth]{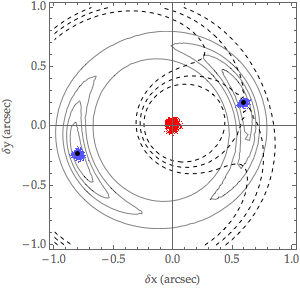}\\
 \includegraphics[width=0.45\textwidth]{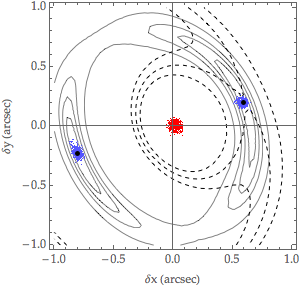}\\
\caption{{Image-plane lens properties of A0140, from SIE models (\textit{top}, b; \textit{bottom}, c). Dashed contours follow the Fermat potential, whereas full lines mark points that would correspond to circular isophotes in the source plane, with radii 0.002$^\ase,$ 0.005$^\ase,$ 0.010$^\ase,$ and 0.025$^\ase.$ Lens models are performed using the constraints from Table~\ref{tab:astromA0140}. The distribution of allowed lens- and image-positions is represented by the swarms, drawn form the lens model (MCMC) posterior.}}
\label{fig:lensA0140}
\end{figure}

Figure~\ref{fig:lensA0140} summarizes the lens model results, displaying the Fermat potential contours (dashed) and a set of isophotes in the hypothesis of a circular source, for best-fitting SIE models (b,c). The inferred parameters are listed in Table~\ref{tab:A0140mods}. The resulting Einstein radius from (a,b) is $\theta_{\rm E}=(0.73\pm0.01)^\ase,$ very close to half the A-B image separation, and slightly higher for models with more substantial shear/ellipticity. The axis ratio from mod.(b,SIE) is $q=0.96\pm0.03,$ significantly rounder than the value suggested by IMACS images. This may mean that either the lens axis ratio from the IMACS image is biased (due to the proximity of quasar images, coarse resolution, PSF mismatch), or that the overall quadrupole is small but substantial shear and ellipticity are present.

 As exemplified in figure~\ref{fig:lensA0140}, models with different shear/ellipticity result in different Einstein ring shapes \citep[see also][]{koc01}. Deeper and higher-resolution imaging data would be useful to: obtain a robust measurement of the lens ellipticity; and detect extended emission from the source host galaxy, thereby adding constraints to the lens model and breaking quadrupole degeneracies.

\begin{table} 
\centering
\begin{tabular}{lccccccl}
\hline
img.	&	$\delta x$ ($^\ase$)	&	$\delta y$ ($^\ase$)	&	mag{\_}i	\\
	&	$=\ -\cos(\mathrm{dec.})\delta$r.a.	&	$\delta $dec.	&	\\
\hline
A & 0.592 & 0.199 & 18.226\\
B & $-$0.804 & $-$0.231 & 18.484\\
G & [ 0.00 & 0.00 ] & 19.253\\
\hline
\end{tabular}
\caption{Astrometry and magnitudes of the three components in A0140. The positions of the two quasar images are given relative to the center of the lens (G). For lens model purposes, we adopt $0.025^\ase$ (i.e. 1px/8) for the uncertainties on positions. The $i-$band magnitudes are calibrated against a bright star in the field of view (r.a.= 01\textit{h}39\textit{m}37.9\textit{s}, dec.= $-$11\textit{d}52\textit{m}37.9\textit{s}, $i=17.67$).}
\label{tab:astromA0140}
\end{table}
\begin{table} 
\centering
\begin{tabular}{lccccl}
\hline
parameter	&	mod.(a)	&	mod.(b)	&	mod.(c)\\
&	(SIS+XS)	&	(SIE)	&	(SIE)\\
\hline
$\theta_{E}$ & $(0.73\pm0.02)^{\ase}$ & $(0.73\pm0.03)^{\ase}$ & $(0.81\pm0.02)^{\ase}$ \\
$q$ & [1.0] & $0.957\pm0.026$ & [0.5] \\
$\varphi_{l}$ & --- & $12.5\pm43.6$ & [28.0] \\
(deg E of N) &  & &  \\
$\varphi_{s}$ & $11.7\pm19.1$ & --- & $-60.7\pm2.1$\\
(deg N of W) &  &  &  \\
$\gamma_{s}$ & $0.013\pm0.008$ & [0.0]& $0.192\pm0.008$\\
\hline
\end{tabular}
\caption{Inferred lens parameters from two different models: (a) a Singular Isothermal Sphere with external shear contributions; (b) a Singular Isothermal Ellipsoid without external shear contributions; and (c) a SIE with $q$ and p.a. as given suggested by the IMACS acquisition image. The angles may vary by $90$~deg depending on the convention chosen for orientations and shear; the Einstein radius is robust against model choice, and the overall quadrupole (shear in mod.a, ellipticity in mod.b) is small.}
\label{tab:A0140mods}
\end{table}

\subsection{Peculiar false positives}
\label{sect:pecfalpos}
Some contaminants, such as single quasars and quasar-star alignments, are common in similar searches, like the SQLS \citep[][]{ogu06,ina12}. Others, such as narrow-line galaxies (NLGs) at $z\approx0.2-0.3$ and star-star alignments, are due to the lack of spectroscopic or UV-excess pre-selection. Some (cf fig.~\ref{fig:others}) deserve a special mention. 
A1507-1442 was flagged as a `sure lens' because visual inspection showed a red galaxy between the two blue point-sources with identical colours identified automatically by cutout modelling, which however the ALFOSC data identified as stars. It had \texttt{ufom}=0.76; its UVx $u-g =+0.49$ (not entering the definition of  \texttt{ufom}) was $\approx1$~mag redder than the threshold $-0.5$ adopted used by \citet{sch16}. In general, this could also indicate a source at $z_{s}\gtrsim2,$ but this was not the case here.
Similar considerations hold for A2243-3840 (\texttt{ufom}=0.56, $u-g = -0.44$).
A2338-2700 shows two blue clumps on either side of a `yellow' galaxy, whose NTT spectra have one secure line (possibly Mg~\textsc{ii} at $z=0.69$) and no information on the central galaxy; preliminary imaging data (SOAR-SAM, PI Motta; not shown here) suggest that this system is probably a merger of NLGs, rather than a low-redshift lens. 
Spectra of A0054-3951 can be clearly de-blended, yielding a galaxy at $z=0.475$ (Mg~\textsc{ii}, O~\textsc{ii}, H$\beta$) or a quasar at $z=1.16$ (C~\textsc{iii}$\left.\right],$ Ne~\textsc{v}, [Ne~\textsc{iv}]), plus a heavily reddened companion.
Finally, the nature of A0008-3655 is unclear, as the lines in the fainter object may just be imperfections in the spectral de-convolution.

\begin{table*} 
\centering
\begin{tabular}{l|d{4.6}d{4.6}c|ccccc}
\hline
name	&	r.a.(J2000)	&	dec.(J2000)	&	mag{\_}i	&	grade	& \texttt{ufom} &	telescope	&	outcome & notes\\
\hline
A1507-1442	 & 	226.9508449	 & 	-14.70332311	 & 	20.10	 & 	2.01	 &	 0.76	&  	NOT	 & contaminant & galaxy seen through two stars\\ 
A1132-0730$^{(s),\ast}$	 & 	173.0309094	 & 	-7.51178122	 & 	18.31	 & 	2.4	 &	0.44	& 	NOT	 &  NIQ &  $z_{s}=1.99$\\ 
A1112-0335$^{\ast}$	 & 	168.1809603	 & 	-3.58591963	 & 	19.33	 & 	2.35	 &	0.46	& 	NOT	 & 	NIQ  & $z_{s}=1.27$\\ 
A1428-0302	 & 	217.4895575	 & 	-3.04143697	 & 	19.70	 & 	2.0	 &	--- 	& 	NOT	 & contaminant &	stars?\\ 
A0326-3122$^{\ast}$	 & 	51.5282775	 & 	-31.38157606	 & 	19.54	 & 	2.0	 &	0.47	& 	NTT	 & 	NIQ & $z_{s}=1.35$\\ 
A2338-2700$^{(s)}$	 & 	354.5270526	 & 	-27.01508058	 & 	18.45	 & 	2.50	 &	---	& 	NTT	 & contaminant & $z=0.68$ (Mg ii), merging NLGs\\ 
A2213-2652$^{(s),\ast}$	 & 	333.4101198	 & 	-26.87418806	 & 	18.10	&	2.24	 &	0.51	& 	NTT	 & 	NIQ & $z_{s}=1.27$\\ 
A0008-3655	 & 	2.1224251	 &	-36.92309206	 &	18.57	 &	2.00	 &	0.14	&	NTT	& unclear &	$z\approx1.715$ qso\\
A0015-1116	 & 	3.9825904	 &	-11.28301617	 &	20.02	 &	2.21	 &	0.34	&	NTT	& contaminant &	star+qso $z\approx1.55$\\
A0054-3951	 & 	13.53023503	 &	-39.86433492	 &	19.40	 &	1.50	 &	0.57	&	NTT	& uncertain &	$z=0.475$ or $z=1.16$\\
A0106-1030	 & 	16.70878412	 &	-10.50974705	 &	20.05	 &	2.00	 &	0.22	&	NTT	& contaminant &	star+qso $z=1.995$\\
A0355-3448	 & 	58.75461712	 &	-34.80051941	 &	19.26	 &	2.18	 &	0.42	&	NTT	& contaminant &	qso+qso, $z=1.19,\ 2.04$\\
A2145-3927	 & 	326.3848895	 &	-39.45890851	 &	19.00	 &	2.36	 &	0.19	&	NTT	& contaminant &	single qso $z\approx0.45$\\
A2243-3840	 & 	340.8657909	 &	-38.66898273	 &	19.69	 &	2.09	 &	0.56	&	NTT	& contaminant &	qso $z\approx2.7$ and galaxy\\
A2356-1213	 & 	359.1233919	 &	-12.22509946	 &	19.31	 &	2.31	 &	0.25	&	NTT	& contaminant &	star+qso $z\approx2$\\
\hline
S1128+2402	 & 	172.0770482	 & 	24.03819957	 & 	17.75	 & 	2.83	 &	---	&	NOT	 & NIQ & $z_{s}=1.608$\\ 
S1030+6055	 & 	157.7161208	 & 	60.93512059	 & 	19.25	 & 	2.89	 &	---	&	NOT	 & contaminant &	$z=1.71$ qso+star\\ 
S1332+3433	 & 	203.1819452	 & 	34.5501693	 & 	18.74	 & 	2.20	 &	---	& 	NOT	 & contaminant &	$z=1.925$ qso+star\\ 
S0332-0021	 & 	53.20211793	 & 	-0.3653620	 & 	19.32	 & 	2.78	 &	---	& 	NOT	 & contminant &	$z=1.71$ qso+qso\\ 
\hline
A0140-1152$^{\ast}$	 & 	25.012499	 & 	-11.871944	 & 	17.63	 & 	---	 &	---	 & 	NTT	 & lens & $z_{s}\approx1.805,$ $z_{l}\approx0.277$\\
A1044-1639$^{(s)}$	 & 	161.195833	 & 	-16.657499	 & 	18.33	 & 	---	 &	---	 & 	NTT	 & contaminant & NLG $z\approx0.3$\\ 
A1020-1002$^{(s)}$	 & 	155.2275	 & 	-10.038888	 & 	18.14	 & 	---	 & 	---	 &	NTT	 &	 pair/NIQ &	$z=2.03$ \\ 
A0054-2404	 & 	13.609166	 & 	-24.077777	 & 	19.72	 & 	---	 &	---	 & 	NTT	 & contaminant &	NLG+qso at $z\approx0.35$?\\ 
A0202-2850	 & 	30.54375	 & 	-28.841388	 & 	19.42	 & 	---	 & 	---	 &	NTT	 & contaminant &	NLG $z\approx0.31$\\ 
A0303-3331$^{(s)}$	 & 	45.89875	 & 	-33.526111	 & 	19.20	 & 	---	 & 	---	 &	NTT	 & contaminant &	NLGs?\\ 
A2201-3613	 & 	330.42125	 & 	-36.216666	 & 	15.61  & 	---	 & 	---	 &	NTT	 & contaminant &	stars\\ 
\hline
A1012-0307$^{(s)}$	 & 	153.066249	 & 	-3.11750	 & 	18.05	 & 	---	 &	---	 & 	NTT	 & 	known lens &  $z_{s}=2.745,$ $f_{r}=3.0(\lambda/9000\rm{\AA})^{-1}$\\ 
A1116-0657	 & 	169.0980709	 & 	-6.96079222		 & 17.25 &	2.42 & --- & --- & known lens & not obs.\\
\hline
\end{tabular}
\caption{Model-accepted candidates from ATLAS DR2 (first sub-list), SDSS control set (second sub-list) and ATLAS-DR3 (third sub-list) targets, that were observed with long-slit spectroscopy at NOT and NTT over 2016. The quoted $i-$band magnitudes are \texttt{AperMag6} for ATLAS and \texttt{model} for SDSS. The DR3 targets were not graded. `NLG' stands for `narrow-line galaxy'. A1116-0657,  given in the last line, was re-discovered during the search in the DR2 footprint and is the known lens, small-separation quad HE1113-0641 \citep[][not re-observed]{bla08}.
$^{(s)}$ Classification of some objects was aided by imaging with the SOAR Adaptive Optics Module (in $z^{\prime}$-band; June, November, December 2016; PI Motta), especially for NLG pairs that could not be otherwise resolved. $^{\ast}$Asterisks mark the NIQs that were targeted independently as described by \citet{sch16}. A0140-1152 is spectroscopically confirmed as a lens with $z_{l}=0.277$ (fig.~\ref{fig:A0140}).}
\label{tab:bigtab}
\end{table*}

\subsection{Further candidates from a search in Gaia, and three additional lenses}
Further development can be brought by the high spatial resolution of Gaia \citep{lin16}, where \citet{fin12} estimated $0.06/\rm{deg}^2$ lensed quasars to be found within a limiting magnitude $G=20$. 
In fact, in the current Gaia-DR1, about $20-30\%$ of known quasar lenses and pairs are recognized as multiple sources with separations $\lesssim8^{\ase},$ suggesting that these systems can be found by selecting quasar-like objects in WISE and then retaining those that correspond to Gaia multiplets. The details of this search are discussed elsewhere \citep{agn17}. When applied to objects with a counterpart in the public ATLAS footprint (i.e. covered in at least one band in DR3), this search recognized three known lenses \citep[LBQS1009-0252, RXJ1131-1231, and W2329-1258:][]{hew94, slu03,sch16}. A fourth lens, HE1113-0641 \citep[][at 11$^h$16$^m$-06$^d$57$^m$ in J2000 system]{bla08}, was re-discovered through the target selection of Sect.~\ref{sect:tarsel}, but due to its very small separation it is not resolved as a multiplet by Gaia-DR1.

To facilitate follow-up, in Table~\ref{tab:morecands} we list additional candidates that could not be followed-up before this paper was completed, due to time and visibility constraints. Besides four identified in DR2 target+candidate selection and one found in DR3 with outlier selection \citep{agn17}, eleven (denoted by `WGAhhmm-ddmm' names) are discovered purely from the Gaia multiplet search\footnote{Search performed in November 2016, candidate selection in February 2017.}. Some of these were identified independently by the search of \citet{sch16}, using ATLAS cutout modelling of WISE-selected objects. Others, having ATLAS coverage in just two bands, could not be selected by the cutout modelling approach, but are found by the Gaia multiplet selection. Most of the candidates have Pan-STARRS1 $grizy$ imaging, which we use to further grade them based on visual inspection.

After this paper was completed with the 2016 campaign results, three candidates of Table~\ref{tab:morecands} (WGA0235-2433, WGA0259-2338, WGA0146-1133) were spectroscopically confirmed as lenses/NIQs in the fall of 2017, by two independent campaigns (NTT, PI Anguita; and WHT, Lemon et al., private comm.). These also fall in the DES footprint, where deeper $grizY$ imaging in good seeing conditions shows red galaxies between the pairs of blue images \citep{wgd17}. As the discussion in Section~\ref{sect:A0140} and by \citep{wgd17} demonstrates, lens models on the discovery images are limited to $\approx10\%$ on the Einstein radii, due to systematics from the quadrupole contributions of shear and ellipticity. Besides that, in order to translate Einstein radii into masses, the redshifts of deflectors are needed. For these three systems, the data collected in 2017 are not deep enough to obtain a deflector redshift.

\section{Summary and Prospects}
A rare-object search, charting the VST-ATLAS public footprint for lensed quasars, has given 7 nearly identical quasar pairs (NIQs), including a new quasar lens, out of a sample of 27 objects observed in 2016. The number of NIQs among followed-up systems could be higher, as two objects are confirmed quasars but currently inconclusive and three additional lenses (tab.~\ref{tab:morecands}) have been confirmed as NIQs/lenses in 2017 (PI Anguita; in prep.) after this paper was completed. This experiment demonstrated that previous searches \citep[e.g. the SQLS, ][]{ogu06,ina12,anu16} can be extended to the regime of patchy waveband coverage and absence of $u-$band or spectroscopic pre-selection, still with moderate ($\approx1^\ase$) seeing and depth ($\approx21$ in $i-$band). Due to the heterogeneous quality of survey data and range of expected image separations, a combination of techniques \citep[][and Section 2 above]{mor04,ofe07,agn15,agn17} has been deployed.

The model-based candidate selection (applied to ATLAS-DR2) seems to give results comparable to the outlier-based target selection (applied to DR3), but is a necessary step to ascertain that the detected point sources have compatible colours and in fact it found lenses \citep{sch16} that were not flagged by the target selection. On the other hand, two systems (A1012-0307, A1020-1002) were identified via target selection, whereas independent cutout modelling had excluded them.
While the lenses discovered in previous campaigns seem biased towards the colours of nearby quasars \citep[see][for a discussion]{wil16}, the ones selected here occupy distinctive regions of optical/IR colour-magnitude diagrams (fig.~\ref{fig:targprop}), including sources at redshift $z_{s}>2.$

\begin{table*} 
\centering
\begin{tabular}{l|d{4.6}d{4.6}c|cccclc}
\hline
name	&	r.a.(J2000)	&	dec.(J2000)	&	mag{\_}i	&	grade	& \texttt{ufom} & release & Pan-STARRS1? & 2017 follow-up$^{(d)}$\\
\hline
A1501-1404	 & 	225.4078925	 & 	-14.072688	 & 	19.37	 & 	2.34	 &	0.13 & DR1+2 & good &	\\ 
A1523-0517	 & 	230.9051605	 & 	-5.284710	 & 	18.78	 & 	1.91	 &	0.35 & DR1+2 & good &	\\ 
A1528-1341	 & 	232.1481305 & -13.690049	 & 	18.44	 & 	---	 &	---	 & DR3	 & good & \\ 
\hline
WGA1122-0529	 & 	230.905161 &	-5.284711	 & 	18.45 & 	---	 &	0.52 & DR3+Gaia & good & \\ 
WGA0336-2406$^{(a)}$	 & 54.20990 & -24.105980 & 	18.72	 & 	---	 &	0.45 & DR3+Gaia & dubious &	\\ 
WGA1149-0747	 & 	173.030909	 & -7.511781 & 17.94	 & 	---  &	0.37 & DR3+Gaia & qso+gal? &	\\ 
WGA0235-2433	 &  38.864257 & -24.553678  & 17.12	 & 	--- & ---	 & DR3+Gaia & likely a lens & lens$^{(d)}$ $z_s=1.43$	\\ 
WGA0259-2338	 &   44.889649 & -23.63383  & 18.41	 & 	--- & ---	 & DR3+Gaia & likely a lens &	lens$^{(d)}$\\ 
WGA0146-1133$^{(a)}$ &   26.636987 & -11.560821  & 17.48	 & 	--- & --- & DR3+Gaia & likely a lens & lens$^{(d)}$	\\ 
WGA0343-3309	 & 	55.923589 &	-33.155475	 & 	18.41	 & 	---	 &	---	 & DR3+Gaia & outside footprint &	\\ 
WGA0030-2326	 & 	7.5009411 &	-23.434479	 & 	19.05	 & 	---	 &	---	 & DR3+Gaia & good & low S/N$^{(d)}$	\\ 
\hline
\hline
WGA1216-1138$^{(c)}$	 & 	184.130790 &	-11.644588	 & 	18.29	 & 	---	 &	--- & DR3+Gaia	&  contaminant &	\\ 
WGA1112-1855$^{(c)}$	 & 	168.222452 &	-18.916569	 & 	19.39	 & --- &	--- & DR3+Gaia & contaminant &	\\
WGA1409-1444$^{(c)}$ & 	212.2502123	& -14.733644	 &	17.77 & ---	 & --- & DR3+Gaia & contaminant &	\\ 
A1201-0324$^{(c)}$	 & 	180.267310	 & 	-3.402352	 & 	19.10	 & 	2.45	 & 0.15	 & DR1+2 & contaminant &	\\ 
A1333-0453$^{(c)}$	 & 	203.261960	 & 	-4.898214	 & 	19.41	 & 	2.34	 &	0.19 & DR1+2 & contaminant &	\\ 
\hline
\end{tabular}
\caption{Selected ATLAS-DR3 candidates, identified with various techniques, not followed up during the 2016 campaigns. The upper part of the table collects objects selected as in the previous Sections; the middle part lists candidates selected among WISE-Gaia multiplets. $^{(a)}$ Some of these had also been flagged independently, through cutout modelling of the full footprint \citep{sch16}. $^{(c)}$ Pan-STARRS1 $grizY$ visual inspection excludes some systems with partial ATLAS-DR3 coverage. $^{(d)}$ After this paper was completed, independent campaigns in 2017 (NTT, PI Anguita; and WHT, Lemon et al., private comm. and in prep.) have targeted some of the systems in this Table. Three are confirmed as NIQs in the spectra, and show a red excess in ATLAS and DES images \citep[shown by][]{wgd17}, confirming them as additional lenses. }
\label{tab:morecands}
\end{table*}

\subsection{Spectral classification of pairs/NIQs and contaminants}
The primary aim of this search was to assemble a comprehensive sample of lenses/NIQs over the ATLAS (publicly accessible, DR3) footprint. From a sample of $\approx10^5$ objects with varying amount of multi-band information, over\footnote{This is an estimate of the effective footprint, which accounts for inhomogeneous waveband coverage of the DR3 footprint as of 2016, when the searches were performed.} $\approx3000$~deg$^2$ and mostly brighter than $i=20,$ a sample of $\approx40$ objects was isolated for follow-up spectroscopy (Tables~3, 4). Given these final numbers, the spectra of all candidates can be visually inspected with ease, the most obvious contaminants excluded, and pairs/NIQs or lenses can be examined individually. 

Upcoming spectroscopic surveys will render this task less immediate. For example, Gaia-DR4 is expected to provide low-resolution optical spectra of all detected sources (down to $G\approx21,$ roughly $i\approx23$), over the whole sky, and the Euclid Wide survey (due 2021-2027) will obtain NIR slitless spectroscopy of objects down to $YJH\approx24$ over $15000$~deg$^2.$ If our ATLAS search is to be rescaled to these expectations, the number of spectra to be examined increases significantly and objective criteria should be devised to discard as many contaminants as possible, while also ensuring a complete selection, and retaining a `manageable' sample for further inspection.

Learning from the sample presented in this paper, the following guidelines can be devised for objective (possibly semi-automatic) spectral classification of candidates. Since we are tasked with classification of spectra, a zeroth-order criterion would be the detection of the same (possibly broad) emission lines in both objects, at the same redshifts $z_{s}\gtrsim0.5,$ within a $\delta z\approx0.005$ measurement accuracy. This immediately eliminates chance alignments of different objects, and red galaxies seen between blue stars. A first criterion is: given two spectra of putative multiple images, can they be fit as a common source spectrum with extinction/microlensing effects? Based on the spectra shown above, this translates into a model spectrum and a prescription for flux-ratios and chromatic effects. The flux-ratio laws explored in this paper are either simple constants, in two wavelength ranges, or power-laws to imitate differential extinction. In order to account for the presence of a possible lens, which can contribute on the redder side of the spectra, the goodness-of-fit can be parameterized by two $\chi^2$ values, one from a model fit below $\lambda\lesssim5500\rm{\AA}$ (where the most prominent emission lines lie, typically), and one fit to $\lambda\gtrsim6000\rm{\AA}.$ A second criterion is a refinement of the first: are the flux-ratios consistent among different lines, or among different ranges on the continua? This would then amount to three $\chi^2$ values overall: one for the continuum-subtracted lines, one on the continua below $\lambda\lesssim5500\rm{\AA},$ and one for $\lambda\gtrsim6000\rm{\AA}.$ The combination of these three criteria, accounting for $10-20\%$ discrepancies in the flux-ratios over lines and continua in different wavelength ranges, excludes all of the contaminants shown in this paper and retains all of the pairs/NIQs/lenses. Allowing for chromatic effects is important in order not to lose veritable lenses (e.g. LBQS1009-0252), and recognize possible lenses whose deflectors contribute to the spectra but are not bright enough to be resolved as separate spectral traces. In presence of spectra with good S/N, one can add a third criterion to classify objects as pairs or NIQs/lenses, by requiring that (once continuum-subtracted) the dispersions on corresponding lines are comparable across the multiple-image spectra. Requiring that the spectra contain broad lines at $z\gtrsim0.5$ and with comparable dispersions (\textit{not} simply the equivalent widths) eliminates most contaminants in the form of binary NLGs and binary quasars.

The criteria listed above can be translated into spectral grades, each corresponding roughly to the likelihood of realizing a NIQ/lens. The use of model $\chi^2$ values provides a smooth grading, and uncertainties in the observed spectra can be translated into data-driven uncertainties in the grades. This procedure in turn enables a quantitative (possibly automatic) evaluation of spectra, while also allowing for some flexibility in candidate ranking. A smooth, data-driven ranking scheme, which also incorporates uncertainties and is based on spectroscopic rather than broad-band information, can be tested on data from different campaigns. Scaling to larger samples, the spectra provided by Gaia-DR4 can be used as a testbed on large datasets, in view of automated and spectroscopic lens searches by the Euclid mission.

\section*{Acknowledgments}
The data presented here were obtained in part with ALFOSC, which is provided by the Instituto de Astrofisica de Andalucia (IAA) under a joint agreement with the University of Copenhagen and NOTSA. This paper includes data gathered with the 6.5 meter Magellan Telescopes located at Las Campanas Observatory, Chile.

TT acknowledges support from NSF through grant AST-1450141, and from the Packard Foundation through a Packard Research Felllowship. VM acknowledges support from Centro de Astrof\'{\i}sica de Valpara\'{\i}so. TA and YA acknowledge support by proyecto FONDECYT 11130630 and by the Ministry for the Economy, Development, and Tourism’s Programa Inicativa Cient\'{i}fica Milenio through grant IC 12009, awarded to The Millennium Institute of Astrophysics (MAS). DM acknowledges financial support from the Instrument center for Danish Astrophysics (IDA). KR is supported by Becas de Doctorado Nacional CONICYT 2017.

AA is grateful to Mike Read for explanations on the ATLAS releases and support with the queries, to Alain Smette and Dominique Sluse for advice on quasar pairs, as well as to Johan Fynbo for sharing part of the NOT guaranteed time in February 2016 for this project.

\appendix
\section{The `locus' of SIE+XS solutions for A0140-1152}
A model with SIE and external shear, when fit to the image configuration and flux ratios of a double, is unconstrained.
Here, we explore the main degeneracies among parameters that result in the same observed constraints. In what follows, we fix the lens p.a. to that measured from the cutouts, i.e. 28~deg E of N, with an uncertainty of 5~deg.


Besides the measurements (image positions, flux ratios), an additional constraint is the presence of only two observed images; this splits the parameter space into different connected components, possibly more than two, i.e. there may be two (separate) connected components within which two images are produced. These connected components are not compact, since there are sequence of parameter choices that converge to solutions where neither two nor four (non-degenerate) images are produced.

The `locus' is the set of points (parameter choices) for which two images are produced and the measured constraints are satisfied.
The projection of the `locus' on different parameters is non-trivial; for example, the Einstein radius is $\approx0.73^\ase$ for most values of the axis ratio, increasing towards $\approx0.8^\ase$ when $q$ approaches 0.35 ($\gamma_{s}\approx0.216$). This explains why $\theta_{E}\approx0.73^\ase$ is robust against the chosen model (`a' or `b' in the main text), unless one posits significant ellipticity in the lens.
When $q>0.9, $ the shear is small ($\gamma_{s}<0.02$) and the solutions are almost independent of it.
Within $0.5\lesssim q\lesssim 0.9,$ there is a simple relation between shear amplitude and flattening:
\begin{equation}
\gamma_{s} \approx -0.39(q - 1.0) - 0.018\ ,
\end{equation}
slightly steeper (resp. shallower) at lower (resp. higher) $q.$ At shear values above 0.02, the `locus' projects almost orthogonally on the shear angle, i.e. $\phi_{s}\approx60$~deg almost independently of the shear amplitude. This is just because shear and ellipticity must compensate to yield a small quadrupole (to produce only two images), and the lens p.a. is kept fixed (within 5~degree tolerance).

\begin{figure}
 \centering
 \includegraphics[width=0.45\textwidth]{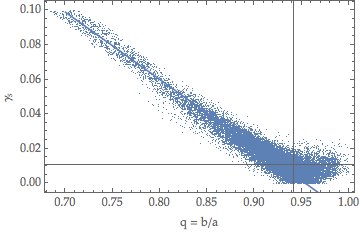}\\
 \includegraphics[width=0.45\textwidth]{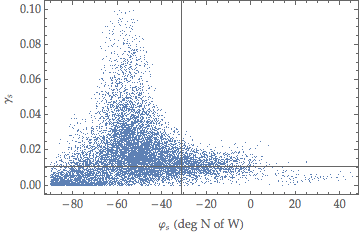}
\caption{{Degeneracies between parameters in mod.(c,SIE+XS), with small uncertaintes on the measured constraints in order to explore a `tube' of general solutions. The lens model parameters are sampled uniformly, except for the lens p.a., which is constrained to its measured value within 5~deg.}}
\label{fig:tube}
\end{figure}

These relations are shown in figure \ref{fig:tube}. Since the `locus' is topologically non-trivial, we explored a tube around it, adopting small but finite uncertainties on the constraints. The non-compactness of the locus can be understood as $\gamma_{s}$ does not approach the same value as $\varphi_{s}$ approaches $\approx60$~deg from the right or from the left.
 The plots are truncated at $\gamma_{s}=0.1,$ just for visual convenience. The shear and flattening `saturate' towards $\gamma\approx0.05$ and $q\approx0.96.$ This also explains the parameters inferred for the restricted models (a,b) in the main text. 
%


\label{lastpage}

\end{document}